%% file: semdate.tex
\documentclass[preprint,5p]{elsarticle}
\usepackage[colorlinks=true,linkcolor=black,anchorcolor=black,citecolor=black,filecolor=black,menucolor=black,runcolor=black,urlcolor=black]{hyperref}
\usepackage{listings}
\usepackage{amsmath,amssymb,amsfonts}
\usepackage{amsthm}
\usepackage{booktabs}
\usepackage[table,xcdraw]{xcolor}
\usepackage{cleveref}
\usepackage{algorithmic}
\usepackage{wrapfig}
\usepackage{graphicx}
\usepackage{textcomp}
\usepackage{xcolor}
\usepackage{adjustbox}
\usepackage{multirow}
\usepackage{adjustbox}
\usepackage{color,soul}
\usepackage[framemethod=TikZ]{mdframed}
\usepackage{scalerel} 

\definecolor{codegreen}{rgb}{0,0.6,0}
\definecolor{codegray}{rgb}{0.5,0.5,0.5}
\definecolor{codepurple}{rgb}{0.58,0,0.82}
\definecolor{backcolour}{rgb}{0.95,0.95,0.92}
\definecolor{light-gray}{gray}{0.80}

\usepackage{makecell}
\usepackage{fontawesome}
\usepackage{tablefootnote}

\newcommand*\rectangled[1]{\tikz[baseline=(char.base)]{\node[shape=rectangle,draw,inner sep=2pt] (char) {#1};}}
\newcommand{\ahref}[2]{\href{#1}{\nolinkurl{#2}}}
\makeatletter
\let\old@lstKV@SwitchCases\lstKV@SwitchCases
\def\lstKV@SwitchCases#1#2#3{}
\makeatother
\usepackage{lstlinebgrd}
\makeatletter
\let\lstKV@SwitchCases\old@lstKV@SwitchCases

\lst@Key{numbers}{none}{%
    \def\lst@PlaceNumber{\lst@linebgrd}%
    \lstKV@SwitchCases{#1}%
    {none:\\%
     left:\def\lst@PlaceNumber{\llap{\normalfont
                \lst@numberstyle{\thelstnumber}\kern\lst@numbersep}\lst@linebgrd}\\%
     right:\def\lst@PlaceNumber{\rlap{\normalfont
                \kern\linewidth \kern\lst@numbersep
                \lst@numberstyle{\thelstnumber}}\lst@linebgrd}%
    }{\PackageError{Listings}{Numbers #1 unknown}\@ehc}}
\makeatother

\lstdefinestyle{mystyle}{
 backgroundcolor=\color{backcolour},
 commentstyle=\color{codegreen},
 keywordstyle=\color{magenta},
 numberstyle=\tiny\color{codegray},
 stringstyle=\color{codepurple},
 basicstyle=\ttfamily\footnotesize,
 breakatwhitespace=false,
 breaklines=true,
 captionpos=b,
 keepspaces=true,
 numbers=left,
 numbersep=5pt,
 showspaces=false,
 showstringspaces=false,
 showtabs=false,
 tabsize=1,
 xleftmargin=2em,
 frame=lines,
 framexleftmargin=1.5em,
}

\theoremstyle{remark}
\newtheorem{rem}{\textit{Definition}}[section]
\usepackage{pifont}
\newcommand{\cmark}{\ding{51}}%
\newcommand{\xmark}{\ding{55}}%

\input{abbreviations}

\begin{document}

\title{Can We Trust Tests To Automate Dependency Updates? A Case Study of Java Projects}

\author[1]{Joseph Hejderup\corref{cor1}}
\ead{j.i.hejderup@tudelft.nl}

\author[1]{Georgios Gousios}
\ead{g.gousios@tudelft.nl}

\cortext[cor1]{Corresponding author}

\affiliation[1]{organization={Delft University of Technology}, 
                addressline={Van Mourik Broekmanweg 6},
                postcode={2628 XE}, city={Delft},
                country={The Netherlands}}

\begin{abstract}
Developers are increasingly using services such as Dependabot to automate
dependency updates. However, recent research has shown that developers perceive such
services as unreliable, as they heavily rely on test coverage to detect
conflicts in updates. To understand the prevalence of tests exercising
dependencies, we calculate the test coverage of direct and indirect uses of
dependencies in 521 well-tested Java projects. We find that tests only cover
58\% of direct and 20\% of transitive dependency calls. By creating 1,122,420
artificial updates with simple faults covering all dependency usages in 262
projects, we measure the effectiveness of test suites in detecting semantic
faults in dependencies; we find that tests can only detect 47\% of direct and
35\% of indirect artificial faults on average. To increase reliability, we
investigate the use of change impact analysis as a means of reducing false
negatives; on average, our tool can uncover 74\% of injected faults in direct
dependencies and 64\% for transitive dependencies, nearly two times more than
test suites. We then apply our tool in 22 real-world dependency updates, where
it identifies three semantically conflicting cases and five cases of unused
dependencies. Our findings indicate that the combination of static and dynamic
analysis should be a requirement for future dependency updating systems.
\end{abstract}

\begin{keyword}
semantic versioning \sep library updates \sep package management \sep dependency management  \sep software migration
\end{keyword}

\maketitle

\section{Introduction}
\begin{figure}[tb]
\includegraphics[width=\columnwidth,keepaspectratio]{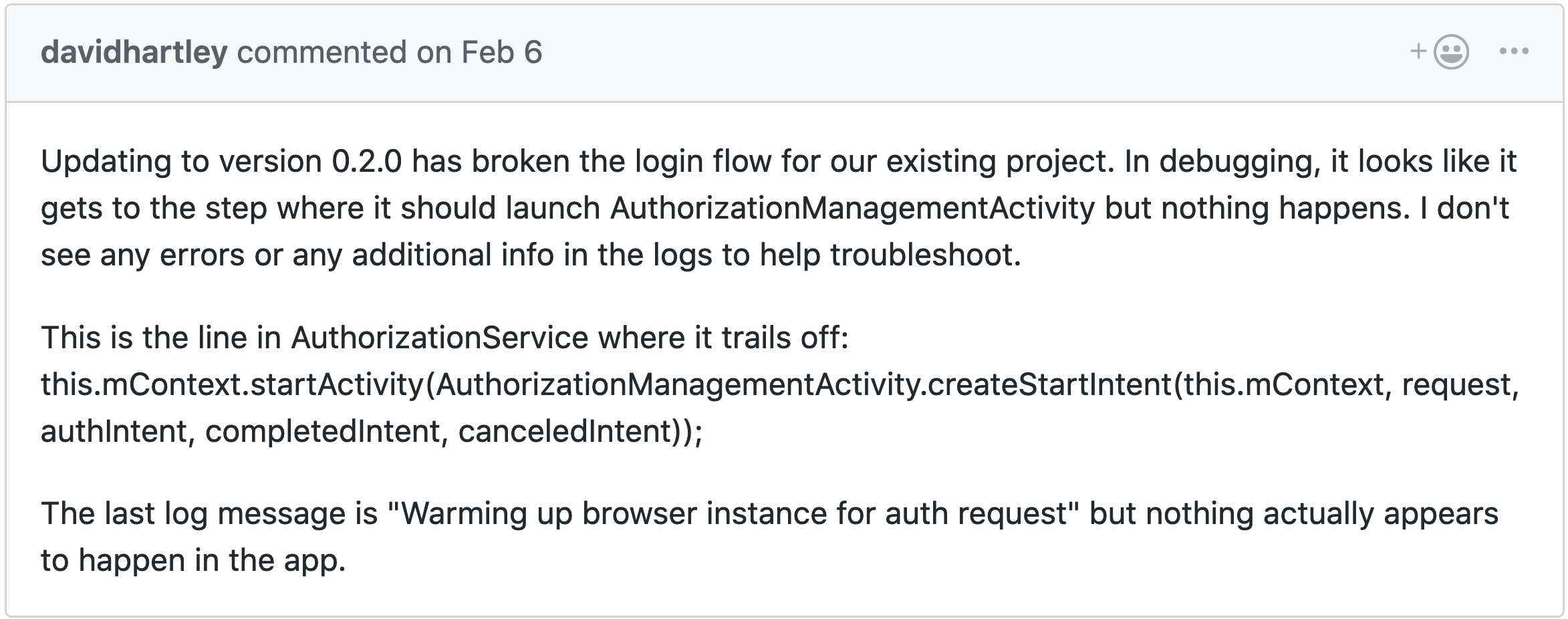}
\caption{\small Update failure in \texttt{okta/okta-sdk-appauth-android}, \#81}
\label{gh:issue}
\vspace*{-1.5em}
\end{figure}

\begin{sloppypar}
Modern package managers facilitate reuse of open source software libraries
by enabling applications to declare them as versioned dependencies.
Crucially, when a new version of a dependency is made available, package
managers will automatically make it available to the client application.
This mechanism helps projects stay up-to-date with upstream developments, such as
performance improvements or bug fixes, with minimal fuss.
Typically, package managers implement a set of interval operators (dependency
version ranges) on top of the SemVer protocol~\cite{npm:guidelines}
that developers use to declare update
constraints. For example, a dependency declared with the range $>= 1.0.0 <
1.5.0$ restricts updates to backward-compatible changes up to $1.5.0$. On the
other hand, $>= 1.0.0$ welcomes automatic updates of all new version releases
starting from $1.0.0$. Given a new library release with version $1.5.0$, the
latter constraint will allow an update but the former will not.
\end{sloppypar}
In practice, most package managers use a liberally
interpreted version of the SemVer protocol with no vetting, allowing library
maintainers to release new changes based on their self-interpretation of backward
compatibility~\cite{npm:guidelines,bogart2016break}. As a consequence, client
programs may unexpectedly discover regression-inducing changes, such as bugs or
semantic changes that break code contracts. Discovering, debugging and
resolving such issues, as exemplified in \Cref{gh:issue}, remains a challenging
task for development teams~\cite{bogart2016break}. In fact, unexpected
regressions are one of the main reasons that deter
developers from upgrading dependencies to new versions~\cite{kula2018developers}.

\begin{sloppypar}
Developers can mitigate the risk of integration errors by either using 
restrictive strategies, such as \textit{version locking}, or permissive strategies
involving \textit{dependency update tooling}. Version locking effectively makes
the dependency tree of client programs immutable and disables automated updates.
This strategy offers maximum stability but is prone to incurring technical debt
due to outdated dependencies. Moreover, developers need to manually discover and
apply security hotfixes. On the other hand, dependency update checkers analyze
version compatibility before deciding to update. There are two main techniques
for deciding version compatibility, \textit{breaking change
detection}~\cite{mezzetti2018type,brito2018apidiff,foo2018efficient} and
\textit{regression testing}~\cite{agrawal1995fault,cleve2005locating}. Detecting
potential breaking changes (i.e., backward API incompatibilities) prevents client
programs from updating to versions that will result in compile failures. A major
shortcoming of this technique is that it depends on the compilation and the
existence of a static type system; many of today's most popular languages are
dynamically typed. A more popular option among developers is the use of services
providing automated dependency updating, such as
\texttt{greenkeeper.io}~\cite{greenkeeper}, \dependabot~\cite{dependabot}, and
\texttt{renovate}~\cite{renovate}, that use project test suites to detect
regression changes on every new update.
\end{sloppypar}

The effectiveness of such services depends highly on the quality of end-users
test suites~\cite{inozemtseva2014coverage}. Poor test coverage of dependency
usage in client code can lead to missing update-induced regressions. Recent
studies~\cite{hilton2018large,kochhar2017code} suggest that high statement
coverage in test suites does not guarantee to find regressions in code changes.
Failing to detect regressions stemming from updates can have dire consequences
for client programs: for example, users dependent on \npm's \texttt{event-stream} package did
not notice a malicious maintainer planting a hidden backdoor for stealing
bitcoin wallets inside the library's source code~\cite{npm:eventstream}.
Moreover, a recent qualitative study~\cite{mirhosseini2017can} also revealed that
developers are generally suspicious of automatically updating their
dependencies. One of the prime reasons is that developers perceive their tests
as unreliable. To reduce the number of false negative updates, we
develop a static change impact analysis for dependency updates called
\uppdatera. By statically identifying changed functions and approximating
call-relationships between an application and its dependencies, change impact
analysis can fill in gaps where test suites have limited coverage or
cannot reach.

In this paper, we set out to empirically understand how reliable developer tests
are in automated dependency updating by addressing the following research
questions: 

\begin{sloppypar}
\begin{itemize}
\item \textbf{RQ1:} \textit{Do tests cover uses of third-party libraries in projects?}
\item \textbf{RQ2:} \textit{How effective are project test suites and change impact analysis in detecting semantic changes in third-party library updates?}
\item \textbf{RQ3:} \textit{How useful is static analysis in complementing tests for compatibility checking of new library versions?}
\end{itemize}
\end{sloppypar}
To study the prevalence of tests exercising dependencies in projects, we first
establish all uses of library functionality from direct and transitive
dependencies in 521 well-tested projects and then measure how much test suites
cover those usages. By
systematically mutating dependency uses in 261 projects, we then conduct a
comparative study on the adequacy of test suites and change impact analysis in
detecting artificial updates with simple faults. To understand the strengths and
weaknesses of using static analysis as a complement to tests in a practical
setting, we evaluate the performance of test suites and \uppdatera on 22 newly
created pull requests that update dependencies.

Our results indicate that tests lack considerable coverage of function calls in
projects that target library dependencies; average coverage is 55\% for direct
dependencies and 26\% for transitive dependencies. Similarly, the average
effectiveness of test suites is 47\% for direct dependencies and 35\% for
transitive dependencies. When using change impact analysis, the average
effectiveness increases to 74\% for direct dependencies and 64\% for transitive
dependencies, suggesting that static analysis can cover open coverage gaps in
tests. Through our manual analysis, we identify that \uppdatera is more
effective in avoiding faulty updates than tests. However, it is prone to false
positives due to difficulties in evaluating over-approximated
execution paths.

Our findings raise awareness of the risks involved with
automated dependency updating. Tool creators should consider
reporting how adequately project tests exercise changed functionality in
libraries under update. In future updating systems, tool creators should
investigate hybrid workflows to complement gaps in regression testing with
static analysis and help developers with prioritizing testing efforts.

\section{Background}

\subsection{Package Managers}\label{sec:repo}
\begin{sloppypar}
Package managers such as Java's \maven, JavaScript's \npm, or Rust's \cargo
provide tooling to simplify the complexities of maintaining, distributing, and
importing external third-party software libraries in development projects. As a
community service to its users, package managers also host a public Online
Package Repository (OPR) where developers can freely contribute with new
packages (e.g., a database driver) or build upon existing
packages (e.g., use a parser library to build a JSON parser). This helps package
manager users to reduce development efforts by benefitting from existing
functionality in their language environments. In a
nutshell, a package is a distributable, versioned software library.
\end{sloppypar}
Because of the relative ease of building packages on top of each other, OPRs
today grow quickly and become evermore
inter-dependent~\cite{decan2017empirical,kikas2017structure}. As a consequence,
package manager users experience a dynamic growth of new hidden dependency
imports in their projects and frequent dependency updates that increase the risk
of build failures due to breaking backward
compatibility~\cite{decan2018empirical,decan2018impact,wittern2016look,bogart2016break}.
The risk of breaking backward compatibility varies between OPRs: \npm and
\mavencentral move the burden of checking incompatible changes on its users
while R/CRAN minimize this risk by requiring a mutual change-cost negotiation
between library maintainers and their users~\cite{bogart2016break}. Users of
OPRs, such as \npm or \maven, either uses additional tooling or disable
dependency updates through version-locking as a protective measure.
Version-locking dependencies guarantee a stable build environment. Additional
tooling provides an extra layer control by scanning dependencies for
vulnerabilities~\cite{decan2018impact}, freshness~\cite{cox2015measuring} or
update compatibility~\cite{mezzetti2018type,foo2018efficient}.

\subsection{Safe Backward Compatible Updates}
\begin{sloppypar}
Update checkers such as \texttt{cargo-crusador}~\cite{cargo:crusador},
\texttt{JAPICC}~\cite{java:japicc}, and \texttt{dont-break}~\cite{npm:dontbreak}
typically determine backward compatibility by ensuring that the new version is consistent
with the public API contract of the old version. Removals or changes in method
signatures, access modifiers, and types (e.g., classes and interfaces) are
examples of inconsistencies that can lead to compile failures in client
code~\cite{dietrich2014broken,raemaekers2017semantic}.
\end{sloppypar}
\begin{lstlisting}[
    language=Java,
    linebackgroundcolor={%
    \color{backcolour}
    \ifnum\value{lstnumber}=19
            \color{green!35}
    \fi
    \ifnum\value{lstnumber}=21
            \color{green!35}
    \fi
    \ifnum\value{lstnumber}=28
    \color{green!35}
    \fi
    \ifnum\value{lstnumber}=31
    \color{green!35}
    \fi
    \ifnum\value{lstnumber}=18
    \color{red!35}
    \fi
    \ifnum\value{lstnumber}=27
            \color{red!35}
    \fi
    \ifnum\value{lstnumber}=30
            \color{red!35}
    \fi
    },
    float=t,
    floatplacement=t!,
    belowskip=-15pt,
    caption=Changes in dependencies that break client semantics,
    label=code:example
    ]
 package client {
     import p2.B;
     class Main {
        void int main {B.b(); B.z();}
     }
 }
 package p2 {
     import p1.A;
     class B {
        int b() {
            int y = 1;
            if A.v(y){y+2;}
            int x = A.a();
            if x > 0 {return 0;}
            return x + y;
        }
        //...
    -   bool z() {return false;}
    +   bool z() {return make_false();}
        //...
    +   bool make_false() {return false;}
    }
 }
package p1 {
    class A {
        //...
    -   int a() {return 0;}
    +   int a() {return 1;}
        //...
    -   bool v(int a) {a > 0 ? true : false}
    +   bool v(int a) {a == 0 ? true : false}
    }
 }
\end{lstlisting}

Checking dependency updates for API inconsistencies is a necessary
precondition to a safe update, but not a sufficient one.
From \Cref{code:example}, we consider an additional class of changes,
\textit{semantic changes}, that are API-compatible (i.e., respects the public
API contract) but introduces incompatible behavior (i.e., regression changes)
for clients after dependency updates. The code example illustrates a
\texttt{client} that depends on \texttt{p2} which in turn depends on
\texttt{p1}. There are two changes that are not semantic preserving in
\texttt{p1}: \texttt{a()} returns 1 instead of 0 (line 27-28) and \texttt{v(int a)}
compares variable \texttt{a} with a different comparison operator (line 30-31).
On the other hand, the change in
\texttt{p2} is semantic preserving: \texttt{z()} still returns \texttt{false}
despite replacing it with a method call to \texttt{make\_false} (line 18-21).
Given a scenario in which \texttt{client} automatically updates to the next
release of \texttt{p1}, and \texttt{p1} updates to the next release of
\texttt{p2}. The changes made in \texttt{p1} will indirectly impact the
behavior of \texttt{client} despite seeming hidden and distant. The change in
\texttt{a()} of \texttt{p1} results in \texttt{b()} to match the if statement on
line 14 and return $0$ instead of doing an addition of \texttt{x} and \texttt{y}
in \texttt{p2} (line 15). This further propagates to the \texttt{client} where
\texttt{b()} is called. Similarly, the change in \texttt{v()} flips the
condition to \texttt{false} instead of \texttt{true} in \texttt{p2} which result
in skipping \texttt{y+2} at line 12. These two code changes illustrate how the
client behavior or the execution flow is not honored after updating to a newer
version.
\begin{sloppypar}
Unlike breaking API contracts, semantic changes are not inherently bad:
the refactoring of \texttt{z()} in \texttt{p2} introduces a new execution path
(e.g., new behavior) to \texttt{make\_false} which continues to return
\texttt{false} after the change. Source code changes that
preserve the same behavior before and after an update are semantic backward
compatible changes. Deciding semantic backward compatibility is also a
contextual problem: Given another client, \texttt{client2} that use the same
dependency \texttt{p2} as \texttt{client} but don't call \texttt{b()} and
\texttt{z()} (line 4). The same update we illustrate for \texttt{client} is semantic
backward compatible for \texttt{client2} as it functions the same way before and
after the update.
\end{sloppypar}
Following the observations in \Cref{code:example}, we denote a semantic backward
compatible update or \textit{safe update} as the following: We denote
$Lib_1,Lib_2 \in Library$ as two versions of the same library and a client $C$
with dependency tree as $T_C = (V,E)$ where $V$ is a set of resolved versioned
libraries used by $C$, and $E$ is the directed dependence between them.
Let $PDG_{TC}$ represent a sound \textit{program-dependence
graph}~\cite{ferrante1987program} of $T_C$ connecting data and control
dependencies between program statements in both client and dependency code. The
transition $[Lib_1 \rightarrow Lib_2]_C$ represents replacing $Lib_1$ with
$Lib_2$ in client $C$. We arrive at the following definition of a safe update:

\begin{rem}
 Given that $Lib_1 \in T_C$ and a request by a package manager to perform
 $[Lib_1 \rightarrow Lib_2]_C$, let $D = Lib_1 \setminus Lib_2$ be a source
 code diff mapping between $Lib_1$ and $Lib_2$, and function $f: D
 \rightarrow Y$ determine semantic compatibility for diff $d_{i} \in D$ in
 client $C$ where $Y \in \{true, false\}$, an automatic update (or safe
 update) can only be made if and only if $\forall d_{i} \in D, f(d_{1})\land
 f(d_{2}) \ldots \land f(d_{n}) = true$ where $i$ varies from $1$ to $n$ and
 $n$ is the cardinality of set $D$.
\end{rem}

\section{Research Questions}
The goal of this paper is to understand how reliable test suites are as a means
to evaluate the compatibility of updated library versions in projects. To that
end, we study a large number of test suites from Maven-based Java projects that
depend on external libraries. 

Bogart et al.~\cite{bogart2016break} report that developers create strategies to
select high-quality libraries based on signals such as active contributors,
project history, and personal trust in project maintainers to reduce the
exposure of unwanted changes. Thus, in our first research question, we
investigate whether testing of third-party libraries is prevalent and a strategy
to minimize the risk of breaking changes: 

\textbf{RQ1}: Do test suites cover the uses of third-party libraries in
projects?

Mirhosseini et al.'s~\cite{mirhosseini2017can} qualitative study suggests that
developers have trust issues with automated updates and perceive tests as
unreliable. A compelling complement to evaluate the effect of dependency changes
is the use of change impact analysis. We set to measure how
capable both test suites and change impact analysis can catch simple semantic
faults in both direct and indirect uses of third-party libraries:

\textbf{RQ2}: How effective are project test suites and change impact analysis
in detecting semantic changes in third-party library updates?

While static analysis can yield higher coverage, it is also more prone to false
positives by classifying safe updates as unsafe. To understand the strengths and
weaknesses of static analysis in a practical environment, we ask:

\textbf{RQ3}: How useful is static analysis in complementing tests for
compatibility checking of new library versions?

We extract a set of real-world update cases from pull requests generated by the
popular service \dependabot and manually investigate the correctness of each
pull request. Then, we analyze each pull request using change impact analysis to
compare the results with the test suite and our ground truth.

\begin{figure*}[tb]
\centering
\includegraphics[width=2\columnwidth,keepaspectratio]{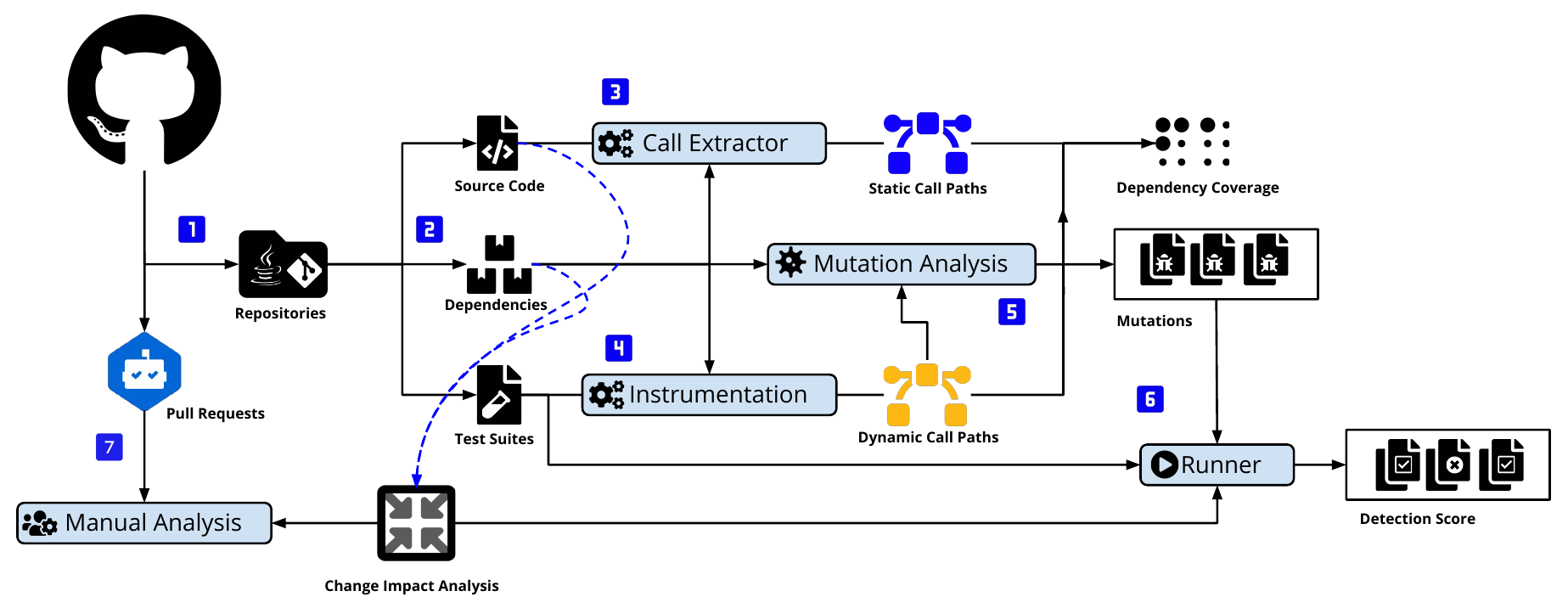}
\caption{Overview of our study infrastructure}
\label{fig:approach}
\vspace*{-1.2em}
\end{figure*}

\section{Research Method}~\label{sec:eval}
We follow the study design depicted in \Cref{fig:approach} to evaluate the
reliability of test suites for automated dependency updating and the potential
of using static analysis. First, we select Java repositories with high-quality
assurance badges and at-least one test class from \github \rectangled{1}.
Then, we build each repository to infer a complete dependency tree of the project
along with its source- and test classes in \rectangled{2}. Second, we feed the
source classes together with the dependencies of a project to the call extractor
and statically extract all its direct and indirect uses of third-party libraries
\rectangled{3}. Third, we use instrumentation to learn all invocations from a
project to its dependencies via its test suite \rectangled{4}. Then, we use the
information from the previous step to calculate the dependency coverage of a
project. Fourth, we generate mutations of dependencies by inserting simple faults
(See~\Cref{tbl:mutants}) in dependency functions executed by tests. Here, we use
dynamic call paths (from \rectangled{4}) to identify such functions \rectangled{5}. We can then run
both the test suite and the change impact analysis to measure the detection
score \rectangled{6}. Finally, we harvest \dependabot pull requests in a
real-time fashion and then manually evaluate how both test suites and change
impact analysis perform in practice \rectangled{7}. 

\subsection{Identifying Usages of Third-Party Libraries}\label{sec:eval:cov}
We refer to the use of third-party libraries as using functionality from
externally-developed libraries in software projects. Specifically, we focus on
functionality exposed as functions in libraries as they are among the most
widespread forms to achieve code reuse. Thus, we consider a function call from a
project to a library dependency as third-party library use. As projects depend
on an ordered tree of library dependencies, there are both implicit and explicit
third-party uses. An explicit use is a direct function call between a project
and one of its declared libraries. On the other hand, implicit use is when a
function in a project transitively calls underlying libraries in a
dependency tree. Given the following example scenario: project \texttt{A}
depends on library \texttt{B}, and library \texttt{B} depends on library
\texttt{C}. If there is a function call path between a function \texttt{a()} in
\texttt{A} to a function \texttt{c()} in \texttt{C} via called functions in
\texttt{B}, project \texttt{A} is implicitly using functionality in library
\texttt{C}.

To identify explicit use of third-party libraries, we statically extract all
function calls to functions that are neither part of the project under analysis or
the Java standard library. By deduction, all such method invocations represent
calls to third-party libraries. For implicit use of third-party libraries, we
statically derive call graphs capturing call paths between a project and its
dependency tree, similar to Ponta et al~\cite{ponta2018beyond}. Finally, we
prune function and call sequences belonging to the Java standard library to
derive a graph representing interactions between a project and its transitive
dependencies.

To measure dependency coverage in a project in \textbf{RQ1}, we use
instrumentation to record all project invocations to third-party libraries
during test suite execution. Using the recorded set, we calculate the proportion
of statically inferred functions covered by the test recorded set of function as
dependency coverage ($Recorded\:functions \subset Declared\:functions$):

\begin{equation*}
Cov_{dep} =\frac{Recorded\:functions}{Declared\:functions}
\end{equation*}

\begin{sloppypar}
Effectively, dependency coverage is function coverage~\cite{myers2011art}, but
only restricted to dependency calls.
\end{sloppypar}
\subsection{Heuristics for Static Impact Analysis}
The central task of automated dependency updating is to facilitate the
continuous integration of new compatible library versions with minimal developer
intervention. Unlike static analysis that may contain
false warnings~\cite{beller2016analyzing}, automated updating suffers instead
from false negatives. A faulty update has a potentially high maintenance penalty
if merged into the project and could cascade into breaking the build of externally
depending projects.

As a step towards reducing false negatives, we are investigating change impact
analysis as a means to potentially reduce coverage gaps where tests are not
able to reach in dependencies. Change impact analysis estimates the reach and
fraction of affected execution paths in a program given a set of code
changes~\cite{arnold1996software}. While there are advancements towards
inference of semantic changes in static analysis such as data flow analysis with
equivalence relations~\cite{gyori2017refining} and mining
techniques~\cite{nguyen2019graph}, precise static interpretation of semantic changes
such as faulty updates is an undecidable
problem~\cite{emanuelsson2008comparative}. Moreover, most of these techniques
only analyze method bodies, and thus not practical for inter-procedural analysis
of projects and their dependencies.

Without possibilities to precisely determine if an update is faulty or not, we
approximate a faulty update (or semantic change) as a change to
the execution flow of a project. We use control flow graphs
(CFGs)~\cite{allen1970control} to represent all possible execution paths of
functions. There are two types of statements in CFG terminology that affect the
execution flow of a program, namely \textit{control} and \textit{write}
statements~\cite{zeller2009programs}. A change to a write statement can affect
the program state (i.e., assign a value to a variable). A change to a control
statement changes the program counter (i.e., determine which statement to be
executed next). By reading the program state, changes to the two other
statements passively impacts \textit{read} statements. Thus, we derive the
following heuristics to classify unsafe updates:
\begin{sloppypar}
\begin{rem}\label{def:change}
Given a diff mapping $D = Lib_1 \setminus Lib_2$ between code entities in
$Lib_1$ and $Lib_2$, we consider a code change as not
\textit{semantic preserving} if and only if $d_i \in D$ has a source
location with a reachable control flow path to client $C$ and maps to the
following potential actions in a CFG:

\begin{enumerate}
\item $d_i$ translates to change in the expression of \textit{write} or
\textit{read} statements (data-flow change)
\item $d_i$ translates to moving a statement from position \textit{x} to
\textit{y} (control-flow change)
\item $d_i$ translates to removing or expanding with new control flow
paths (control-flow change)
\item $d_i$ translates to changes in branch conditions (control-flow change)
\end{enumerate}
\end{rem}
\end{sloppypar}

The definition is an over-approximation; code changes such as refactoring would
result in being classified as an unsafe update if and only if affected functions are
reachable. As services such as \dependabot present only the outcome of test
results and a changelog between the old and new version of a library, change
impact analysis instead precisely pinpoint affected execution paths in an
update. Such information help project maintainers prioritize testing efforts or
determine the potential risk of the update. 

\begin{table}[tb]
\caption{Mutation operators (based on Papadakis et al~\cite{papadakis2019mutation})}
\label{tbl:mutants}
\begin{adjustbox}{width={\columnwidth},totalheight={\columnwidth},keepaspectratio}
\begin{tabular}{@{}lll@{}}
\toprule
Names & Description & Example \\ \midrule
ABS & Absolute Value Insertion & $v \longmapsto \texttt{\textbf{abs}}(v)~|~\texttt{\textbf{-abs}}(v)~|~0$ \\
AOR & Arithmetic Operator Replacement & $x~\texttt{op}~y \longmapsto x~\texttt{\textbf{+}}~y~|~x~\texttt{\textbf{\%}}~y~|~x~\texttt{\textbf{/}}~y$ \\
LCR & Logical Connector Replacement & $x~\texttt{op}~y \longmapsto x~\texttt{\textbf{||}}~y~|~x~\texttt{\textbf{\&\&}}~y~|~x~\texttt{\textbf{\^}}~y$ \\
ROR & Relational Operator Replacement & $x~\texttt{op}~y \longmapsto x~\texttt{\textbf{>}}~y~|~x~\texttt{\textbf{!=}}~y~|~x~\texttt{\textbf{>=}}~y$\\
UOI & Unary Operator Insertion & $v \longmapsto v\texttt{++},\texttt{++}v,\texttt{!}v$ \\ \bottomrule
\end{tabular}
\end{adjustbox}
\end{table}

\subsection{Creating Unsafe Updates in Project Dependencies}
For seamless integration, it is important for automated dependency updating to
detect incompatibilities that arise when updating a library dependency. By using
mutation analysis to seed artificial faults in all uses of third-party libraries
in a project, we can derive an adequacy test of detecting incompatibilities in
automated dependency updates. We first dynamically extract a set of called
third-party functions in a project and then apply mutation operators defined in
\Cref{tbl:mutants} to construct a set of artificial updates that are false
negatives. As static analysis can over-approximate execution paths (i.e., risk
creating false-positive cases), we resort to dynamic analysis to ensure mutations
of truly invoked functions. For the selection of mutation operators, we choose
operators common in mutation testing
studies~\cite{just2014mutants,papadakis2019mutation} that focus on simple
logical flaws and exclude mutation operators with a limited effect such as
deleting statements~\cite{just2014mutants}. 

\begin{sloppypar}
In comparison to using actual update cases, the mutation setup provides a
systematic way to introduce simple faults in \textbf{all} uses of third-party
libraries in a project to measure the effectiveness of detecting faulty updates.
Manually curating false-negative cases of dependency updates limits to specific
project-library pairs and may not generalize to other projects that use the same
library. Moreover, finding such pairs for all libraries in a project to create
an overall assessment may not be possible in most projects.
\end{sloppypar}
For \textbf{RQ2}, we denote \textit{mutation detection score for dependencies}
(an adaption of \textit{mutation score}~\cite{just2014major}) as a tool's
ability to detect a mutated reachable dependency function as (mutants):

\begin{equation*}
Detection\:Score=\frac{Detected\:mutants}{All\:mutants}
\end{equation*}

\subsection{Manual Analysis of Pull Requests}
As the artificially created updates address only false negatives, we also need
to understand how static analysis performs in practice. Thus, we manually
analyze the applicability of static analysis using pull requests through a
lightweight code review. Due to the absence of established ground truth or a
benchmark, we resort to manually creating a ground truth of libraries under
update. As understanding the use context of a project-library is challenging, we
also, attempt to corroborate our findings by posting our assessment as pull
request comments. Below, we define our setup for the manual analysis:

\begin{sloppypar}
\paragraph{Selection criteria} We select pull requests generated from the
popular service \dependabot on \github that supports automated updates of Java
projects using the Maven-build system. To select significant and high impactful
projects and increase the chance for a response by a project maintainer, we
harvest newly created pull requests using \ghtorrent's event
stream~\cite{gousios2012ghtorrent} and adopt the following filter criteria: (1)
\textit{high stargazer, watchers or forks count} indicate popularity, (2)
\textit{no passive users} indicate projects that assign reviewers and frequently
merge \dependabot-pull requests, (3) \textit{dependency type} indicates that we
only consider \maven compile and runtime dependencies, and (4) \textit{project
buildability} indicates that we can compile the project out of the box. 
\end{sloppypar}
\paragraph{Code review protocol}
After a pull request meets the selection criteria, we first inspect the diff in
the pull request to identify the old and the new version number of the library
under update. Then, we download the source jar of the old and new version from
Maven Central and use a diffing tool to localize the set of changes. By
reviewing the change location, consulting the changelog, inspecting the tests of
the library, we classify the nature of a change as refactoring, structural
(i.e., breaking change), or behavioral (i.e., semantic change). Next, we check
out the project at the commit described in the pull request and manually
localize uses of the library by first performing keyword search of import
statements leading to the library under update. Then, we track the data- and
control-flow of imported items (e.g., object instantiations, function
invocations, and interface implementations) to map out how the project uses the
library under update. If the library under update is a transitive dependency, we
first trace how the project uses its direct dependency and then how the used
subset of the direct dependency uses the transitive dependency. After mapping
out uses of the library under analysis, we can then establish whether a project directly or indirectly uses
any of the changed classes and function signatures identified in the diff and
whether those changes make the update safe or not. If the changes do not alter the
logic (e.g., refactorings) of the project, we consider the update safe.
Refactorings are in some cases highly contextual and can yield different
outcomes as exemplified in the following: the changed function~\texttt{foo(x)}
adds a new IF-statement with the condition~\texttt{x > 50} that breaks the
original functionality. Project~\texttt{A} uses~\texttt{foo(x)} indirectly, and
through the manual analysis (including inspection of its tests), we can
establish that
the threshold is $x < 20$ in all cases, and thus the update is safe to make. On
the other hand, project~\texttt{B} has a public function~\texttt{bar(x)} that
passes \texttt{x} in a function call to~\texttt{foo(x)}. Here, we cannot assume
anything around \texttt{x} as users of~\texttt{B} could call~\texttt{bar(x)}
with any \texttt{x}. In this case, we consider the update unsafe.

After manually evaluating pull requests, we classify them using one of three categories:
\begin{itemize}
    \item~\textit{Safe}: the update is safe to perform and will not negatively impact the functionality of the project.
    \item~\textit{Unsafe}: the update is risky and could lead to potential unexpected runtime changes.
    \item~\textit{Unused}: the update of an unused dependency (i.e., it is only declared in the project but not used).
\end{itemize}

Based on the outcome of the update tooling, we compare it with the
classification above and consider the following:

\begin{itemize}
    \item False Negative (FN) when classifying an unsafe update as safe.
    \item False Positive (FP) when classifying a safe update as
    unsafe or falsely updating an unused dependency.
    \item True Positive (TP) when both our manual classification and update
    tooling has the same conclusion.
    \item True Negative (TN) when not creating an update for an unused dependency. 
\end{itemize}

\subsection{Dataset Construction}

\begin{table}[tb]
\caption{Descriptive Statistics for 521 \github projects (each variable aggregated per project)}
\label{tab:desc}
\begin{adjustbox}{width={\columnwidth},totalheight={\columnwidth},keepaspectratio}
\begin{tabular}{@{}lcrrrrc@{}}
\toprule
Variable                       & Unit  & $Q_{0.05}$ & Mean & Median & $Q_{0.95}$  & Histogram \\ \midrule
Project Methods                & count & 20    & 668  & 210.5  & 2320.5 & \parbox[c]{2em}{
    \includegraphics[width=2em, height=2em,keepaspectratio]{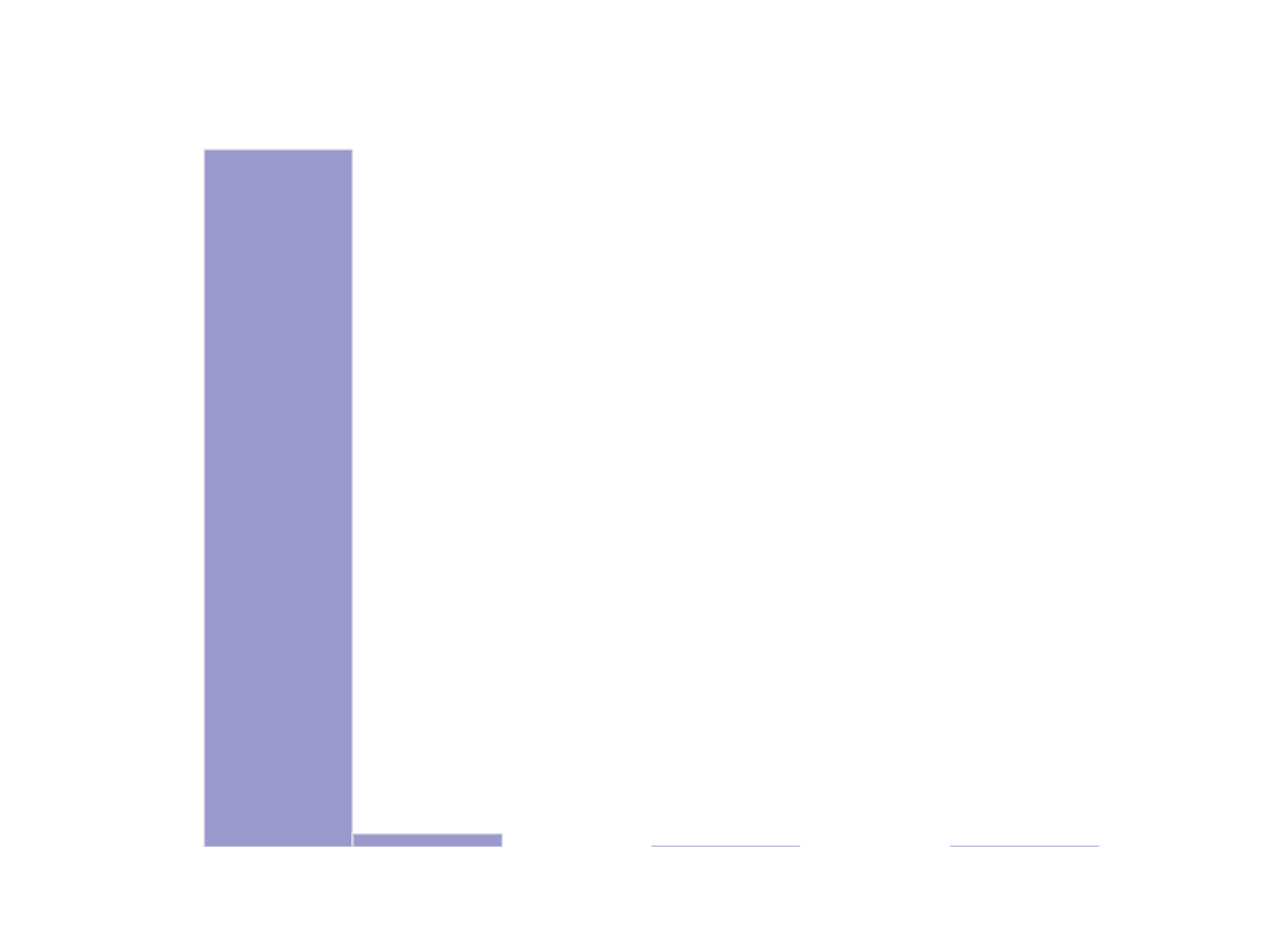}}           \\
Test Coverage (function calls) & \%    & 7     & 72   & 64     & 97     &   \parbox[c]{2em}{
        \includegraphics[width=2em, height=2em,keepaspectratio]{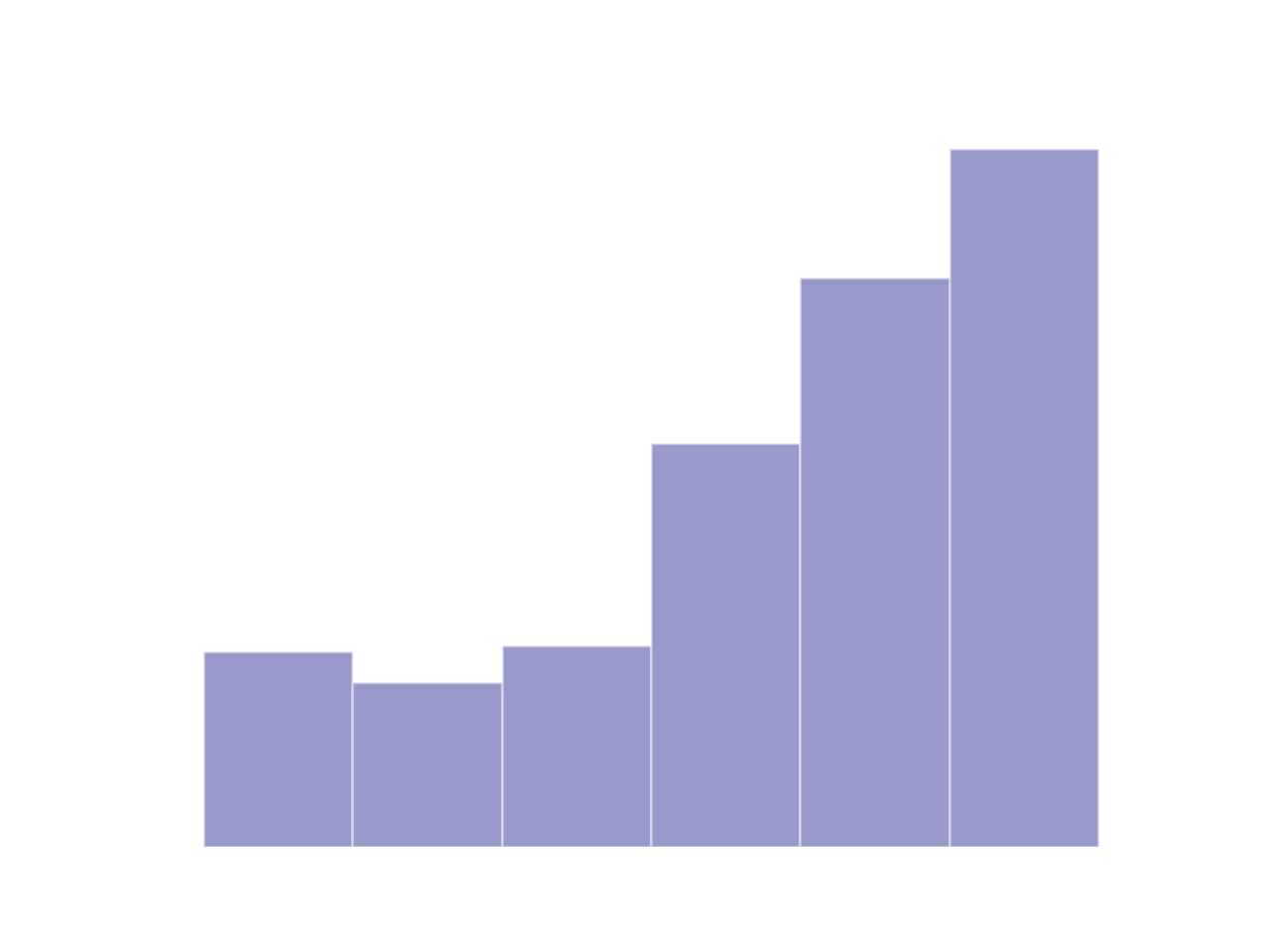}}          \\
Direct Dependencies            & count & 1     & 10   & 7      & 31     &  \parbox[c]{2em}{
    \includegraphics[width=2em, height=2em,keepaspectratio]{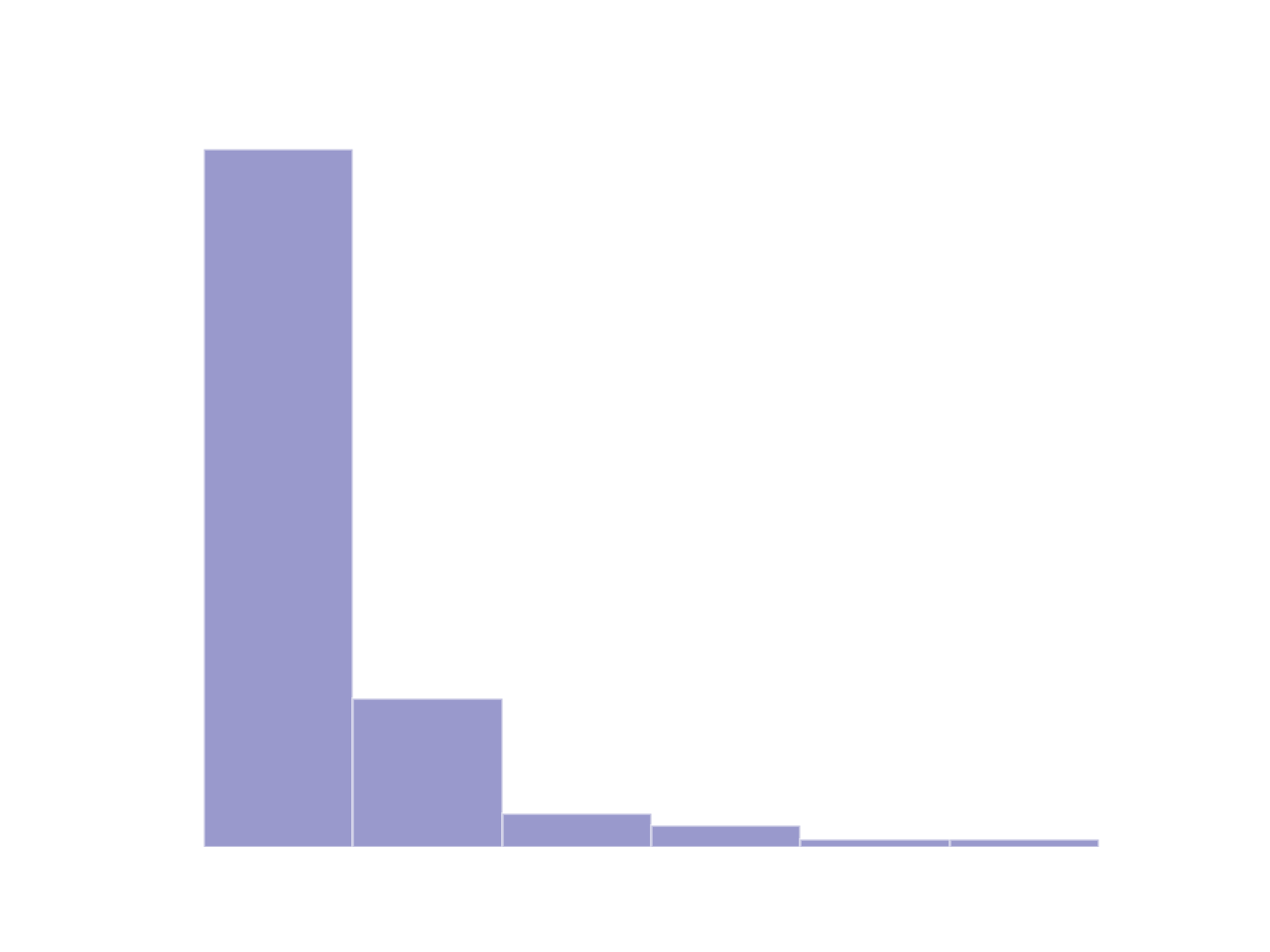}}         \\
Transitive Dependencies        & count & 1     & 31   & 16     & 105    &  \parbox[c]{2em}{
    \includegraphics[width=2em, height=2em,keepaspectratio]{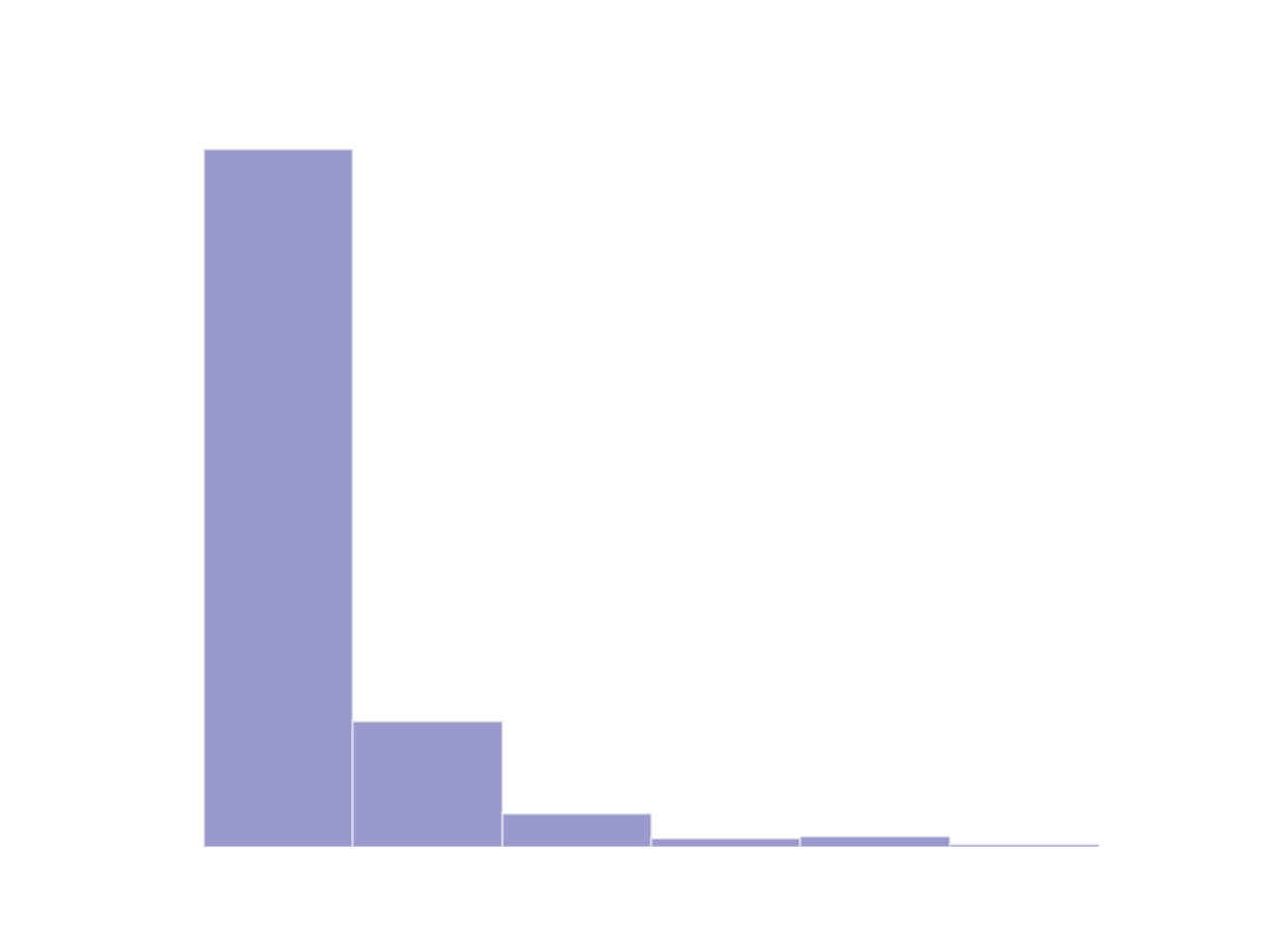}}          \\ \bottomrule
\end{tabular}
\end{adjustbox}
\end{table}

We sample $1,823$ repositories from \github that have Java as the primary
language, \maven as the primary build system, and have at-least a high-quality
assurance badge (i.e., \travis, CodeClimate, coveralls, and CodeCov) as a signal
for having tests~\cite{trockman2018adding}. Services such as \dependabot can
update dependencies in projects as long as there is a valid \texttt{pom.xml}
file. Next, we build and then dry-run projects on both the instrumentation  and
mutation pipeline to eliminate incompatible projects. In total, there are $818$
repositories that compile to Java 8 bytecode and have at least one compiled test
class. Out of the $818$ built projects, $521$ projects successfully run the
instrumentation pipeline, and a subset of $262$ projects are compatible with the
mutation pipeline. The number of projects in the mutation pipeline is nearly
double the ratio of a recent previous study~\cite{zhang2018predictive}.
\Cref{tab:desc} presents descriptive statistics on four aggregated variables for
projects belonging to the instrumentation pipeline. The median number of
declared methods is 210 (mean: 668) with a heavily positive skewed distribution.
75\% of all projects in our sample cluster around $588$ or less declared methods
with $36$ projects having more than $1400$ methods. The largest project is
\texttt{oracle/oci-java-sdk} with $22,264$ methods. As per \Cref{sec:eval:cov},
we measure test coverage of all function calls made in a project. We can observe
that the test coverage is generally high: half of the projects have coverage of
67\% or more. For the number of dependencies, we can observe that the
distribution does not drastically change: the median changes from 7 to 16,
indicating a small expansion of transitive dependencies. Overall, our dataset
represents mid-sized projects that use a significant number of dependencies with
varying test coverage.

\subsection{Implementation}
We discuss the implementation of \uppdatera, a tooling for performing change impact
analysis of library dependencies in \maven, and our pipeline to run our
experiments. We have open-sourced the tooling and docker images for automation and
reproducibility of our study (see \Cref{sec:threats}).

\begin{figure}[tb]
\centering
\fbox{\includegraphics[width=0.47\textwidth,keepaspectratio]{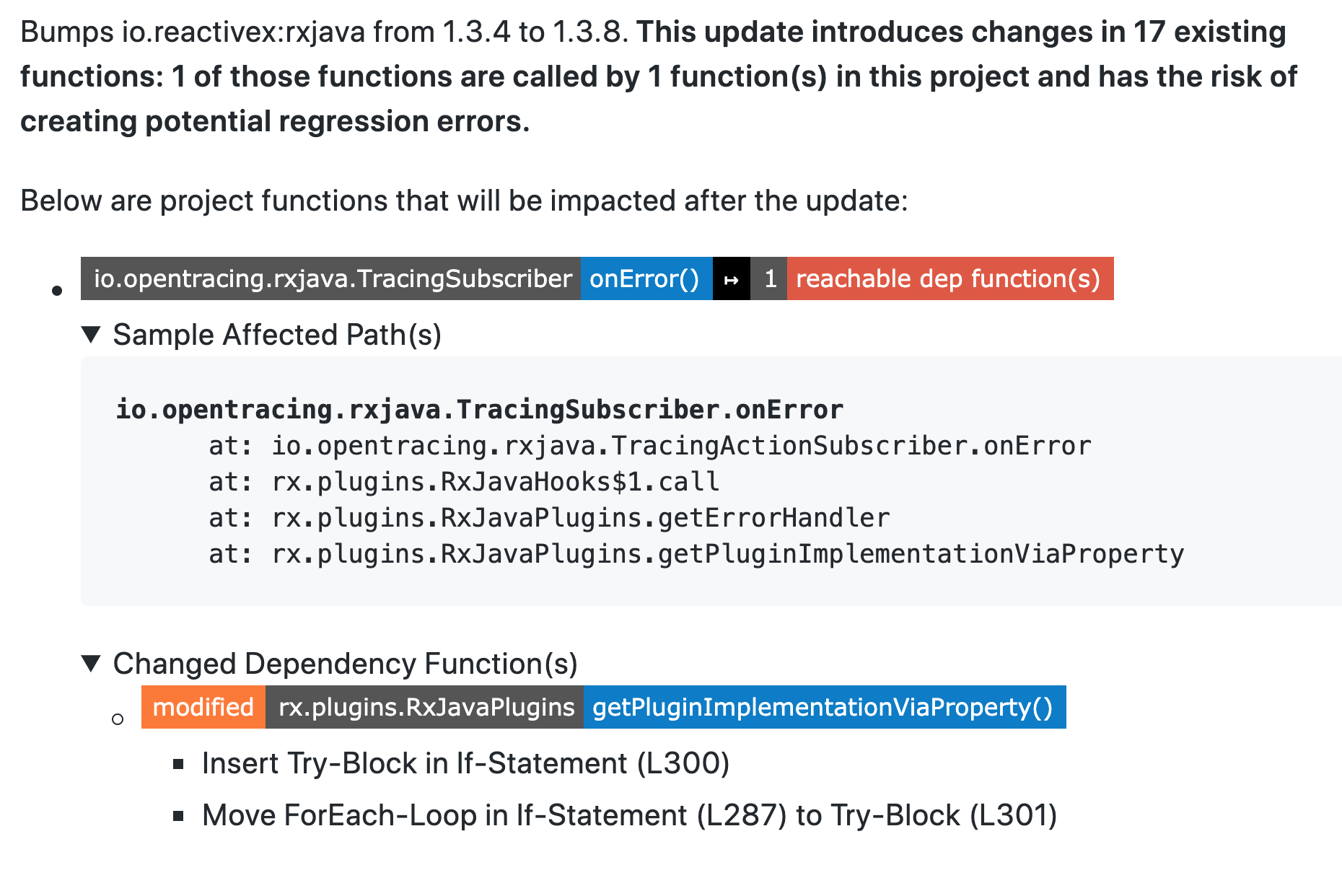}} \caption{Example of updating
\texttt{rxjava} from 1.3.4 to 1.3.8 in the project
\texttt{opentracing-contrib/java-rxjava}}
\label{fig:example}
\end{figure}

\subsubsection{\uppdatera}

\begin{sloppypar}
Given a request to update a dependency to a new version in a \texttt{pom.xml}
file, \uppdatera first performs AST differencing of the current and new version
of the dependency to identify a list of functions with potential behavioral
changes using~\texttt{SpoonLabs/GumTree}~\cite{falleri2014fine}. Then,
\uppdatera computes a call graph inferring all control-flow paths between client
and dependency functions following Ponta et al.~\cite{ponta2018beyond} approach
for call graph construction (using WALA). Finally, \uppdatera performs a
reachability analysis using the list of possible behavioral changes on the call
graph to find reachable paths to the client code. \Cref{fig:example}
demonstrates an example of using \uppdatera for updating the library
\texttt{io.reactivex:rxjava} from version 1.3.4 to 1.3.8 in
\texttt{opentracing-contrib/java-rxjava}. The report features a call stack to
the changed function along with a set of AST diffs. In this particular case, the
\texttt{onError()} function in the class \texttt{TracingSubscriber} transitively
calls \mbox{\texttt{getPluginImplementationViaProperty()}} in the dependency
class \texttt{RxJavaPlugins}. The addition of a \texttt{try-catch} block in the
function takes care of unhandled exceptions which may have been handled by
clients in previous versions (i.e., potential regression change)
\end{sloppypar}

\begin{sloppypar}
In the following, we motivate our implementation choices for a change impact
analysis tool designated for updating library dependencies.
\end{sloppypar}
\paragraph{{\normalfont \textbf{Diffing}}}
\uppdatera performs source code differencing at the abstract syntax tree (AST)
level of both the current and the new version of a dependency to identify
functions with code changes. AST differencing
algorithms~\cite{fluri2007change,falleri2014fine} produce fine-grained and
accurate information about the type and structure of source code changes.
Following \textit{Definition}~\ref{def:change}, we capture AST transformations
at the statement level and map the following as regression changes:
\begin{sloppypar}
\begin{itemize}
 \item Any method-level \textit{move} operation mirrors moving a statement from
 line \textit{x} to \textit{y}.
 \item \textit{deletion}, \textit{update} or \textit{insertion} of
 \textit{Expression} ASTs mirrors data-flow changes.
 \item \textit{deletion}, \textit{update} or \textit{insertion} of control struct ASTs
 such as \textit{IF}, \textit{While}, \textit{FOR} mirrors control-flow changes.
 \item \textit{deletion}, \textit{update} or \textit{insertion} of
 \textit{Call-Expression} ASTs represents changes mirrors control-flow changes.
\end{itemize}
\end{sloppypar}
As an alternative to AST differencing, we could consider bytecode differencing.
Bytecode (e.g., LLVM's IR or JVM code) differencing compute edit scripts at the
instruction level. Although this technique offers a fine-grained and a
compelling alternative to AST differencing, instruction-level changes can be
difficult to understand for developers not familiar with low-level details.

\begin{sloppypar}
\paragraph{{\normalfont \textbf{Call Graph Construction}}}
\uppdatera constructs a call graph capturing inter-procedural control-flow paths
between client and dependency functions. Each node in the call graph represents a
fully resolved function identifier and should be identical to the identifiers in
the changeset of the \textbf{Diffing} phase.

We advocate the use of call graph algorithms that are both
\textit{soundy}~\cite{livshits2015defense} and scalable for analyzing projects
in the wild as a general guideline. The call graph algorithm should
support and resolve as many language features as possible. Limited support of
language features could potentially leave gaps in the coverage of projects
making use of unsupported features. Similar to static analyses of security
applications, achieving high recall is more crucial than precision to avoid
recommending faulty updates.

As recent studies~\cite{kikas2017structure,decan2018empirical} suggest that
irrespective of the OPR, the majority of packages have a small number of direct
dependencies, but a high and growing number of transitive dependencies.  For
example, 50\% of all packages in \crates have a dependency tree depth of at
least 6~\cite{decan2018empirical}. Therefore, performing static analysis at the
boundary of a project and its dependency tree can become computationally expensive
and impractical in DevOps environments. Moreover, as \uppdatera can expect to
analyze any compatible project in the wild, the algorithm should be scalable to
cater large projects and cheap to construct to cut down computation time.  

Finally, a potential trade-off of using call graphs instead of CFGs is the loss of
analysis precision due to the absence of data-flow paths in the graph. However,
taking into account program features such as aliases, arrays, structs, and class
objects in dataflow analysis adds additional complexity and scalability problems
when moving the analysis boundary to include project dependencies. Supporting
such analysis adds extra precision but may not yield extra actionability. 
\end{sloppypar}

\paragraph{{\normalfont \textbf{Reachability Analysis}}}
For each changed function identified in the \textbf{Diffing} phase, \uppdatera
performs a reachability analysis on the call graph to detect paths connecting
changed dependency functions to functions in the analyzed project. If \uppdatera
finds such paths, it marks the update as potentially \textit{unsafe}. If
no such paths are found, \uppdatera marks it as a potential \textit{safe} update
and recommends the update to the package manager. Finally,
\uppdatera also reports the impacted paths between dependencies and project
functions, to inform developers of the program paths that need to be inspected
in response to an update in a dependency. 

\subsubsection{Experimental Pipeline}
\begin{sloppypar}
    
To implement our methodology, we first develop a call extractor that records
complete call sequences between a project and its library dependencies. The
implementation builds on instrumenting library classes using ASM~\cite{java:asm}
and the Maven Dependency Plugin. To infer function calls to libraries from a
project (\textbf{RQ1}), we use ASM to statically extract call sites for direct
dependencies. We generate call graphs using WALA~\cite{ibm2006tj} configured for
the CHA algorithm for transitive dependencies. Following Reif et al.~\cite{reif2019judge}'s
comprehensive benchmark of call graph algorithms for Java, we find that the CHA
algorithm supports the most language features and has a lower runtime in
comparison to more precise points-to analysis algorithms such as 0-1-CFA or
N-CFA.
\end{sloppypar}

For \textbf{RQ2}, we
implement the update emulation pipeline (i.e. mutation analysis) on top of
PITest~\cite{coles2016pit}, a popular in-memory-based mutation testing framework
that works with the popular test runners \junit and \testng by limiting
mutations to library functions identified from the call extractor. We exclude
the use of experimental mutation operators that cannot guarantee non-equivalent
mutations. For each mutated class, we use Procyon~\cite{strobel2016procyon} to
decompile into a source file for AST diffing in the case of \uppdatera.

\section{Results}
Here, we report the results of our research questions.

\subsection{RQ1: Dependency coverage}

\begin{figure}[tb]
\centering
\includegraphics[width=0.47\textwidth, keepaspectratio]{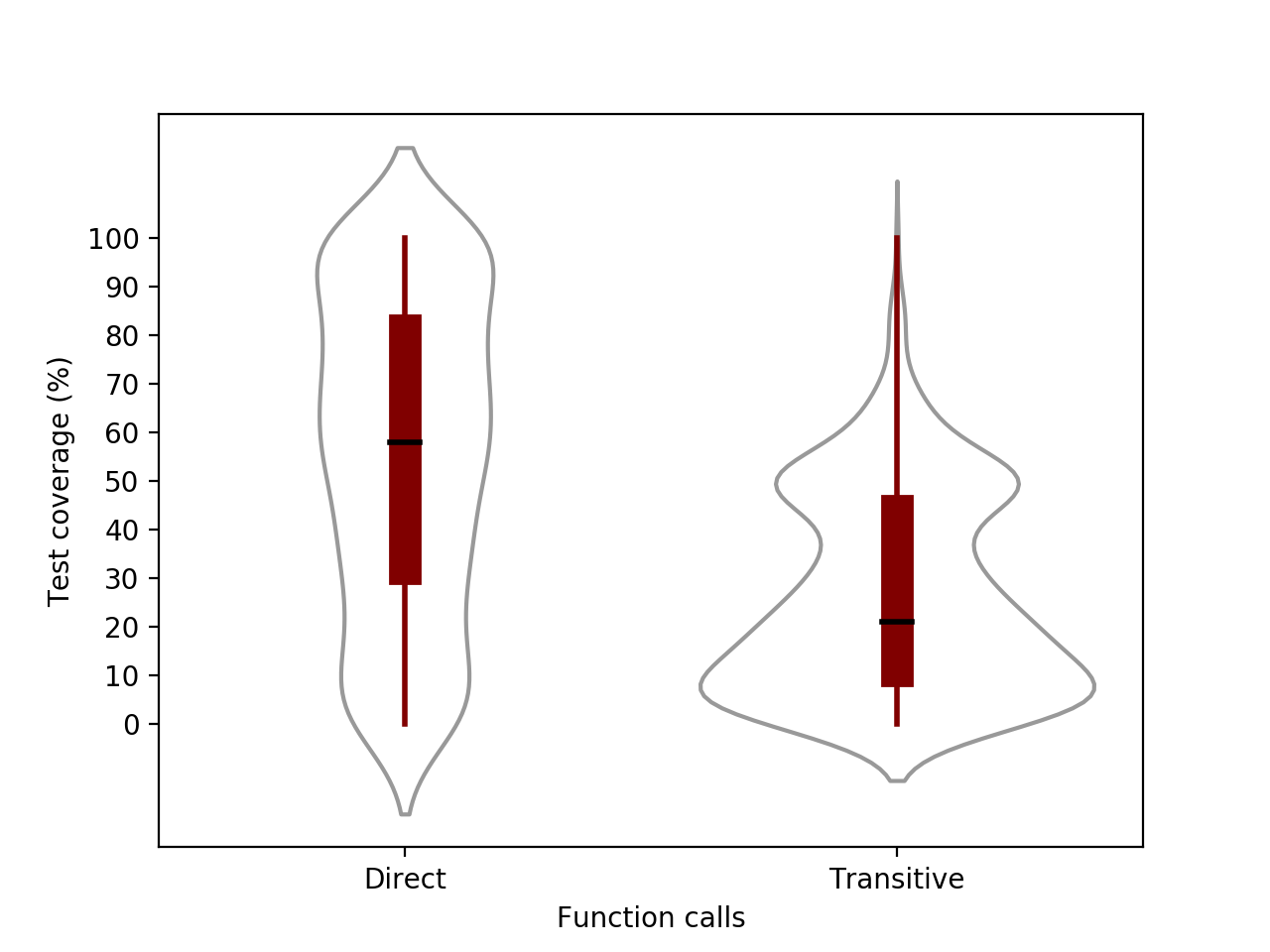}
\caption{Test coverage of dependencies}
\label{fig:cov}
\end{figure}

\Cref{fig:cov} presents a violin plot of dependency  coverage on the left-hand
side, and dependency coverage including transitive dependencies on the
right-hand side. Overall, 13\% (67/521) projects have less than 10\% coverage,
suggesting at large that a majority of projects have some tests exercising at
least one dependency use. We observe that the median coverage is 58\% (mean:
55\%): half of the \github projects miss coverage of more or at least 42\% of
all dependency function calls. In practice, there is a risk that automated
dependency updating may not have tests that exercise changes in dependencies.

The right-hand side of \Cref{fig:cov} shows the dependency coverage taking into
account reachable paths to transitive dependencies in projects. The distribution
has a bimodal shape with two peaks at, 9\%, and at 52\%, suggesting two classes
of projects. In the first class, half of the projects have a median dependency
coverage of 21\% (mean: 26\%), indicating that project test suites at large do
not exercise dependencies in depth. This is not surprising: an ergonomic factor
of third-party libraries is that they are well-tested and should in principle
not need extra tests~\cite{cox2019surviving}. In the second class, we can
observe that projects have tests that exercise dependencies in-depth, suggesting
the presence of projects with adequate test suites. As mentioned in
\Cref{sec:eval:cov}, these results are indicative as we compare against
statically inferred call paths, which, being over-approximating, may not be
representative of actual calls.

\begin{mdframed}[roundcorner=5pt,nobreak=true,align=center]
 \textbf{Findings from RQ1:}\textit{
   Half of the 521 projects exercise less than 60\% of all direct dependency calls
   from their tests; this drops to 20\% if paths to transitive dependencies are
   considered.
}
\end{mdframed}

\begin{figure}[tb]
\centering
\includegraphics[width=0.47\textwidth,keepaspectratio]{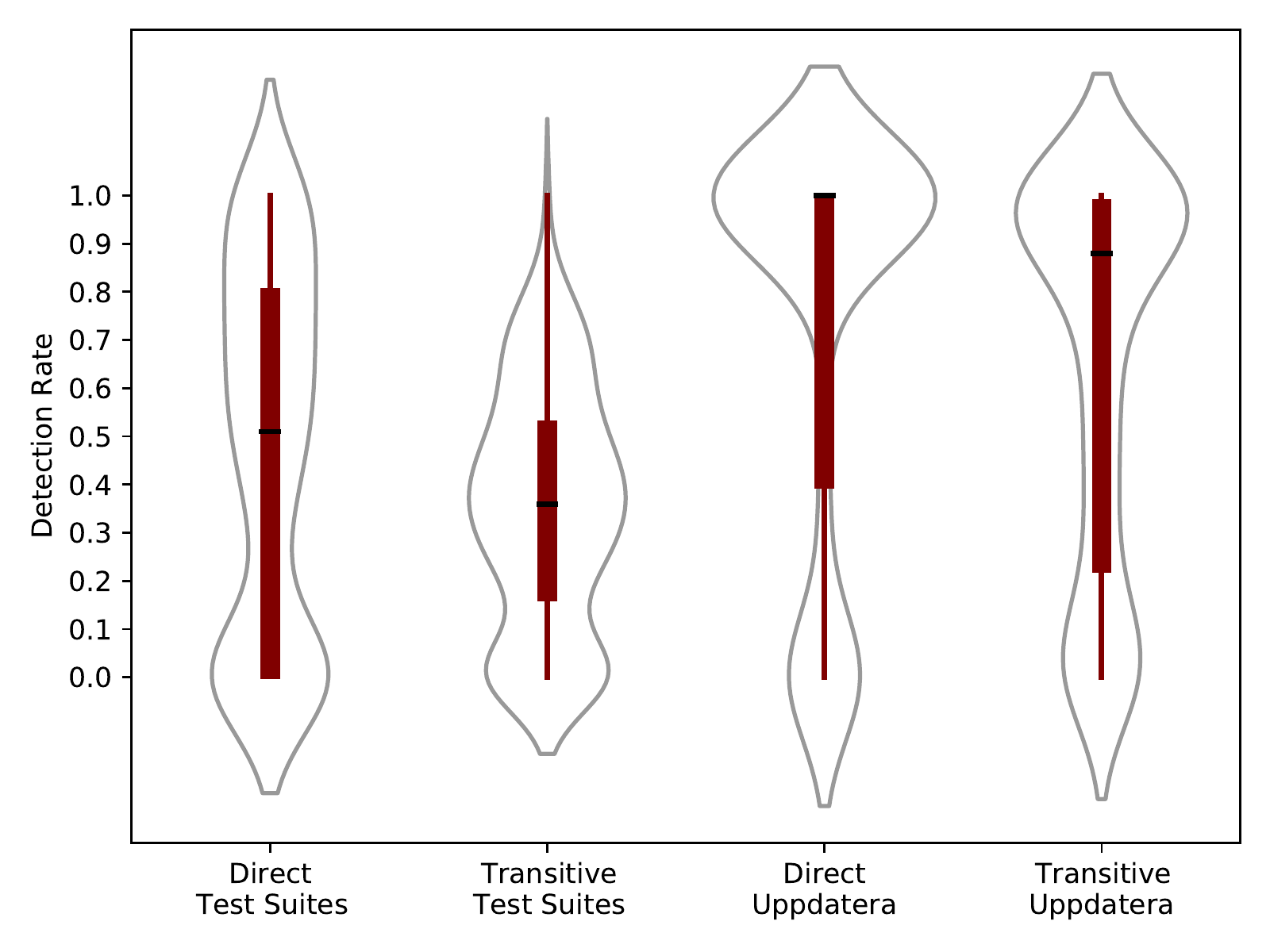}
\caption{Mutation detection score}
\label{fig:regr:updates}
\end{figure}

\subsection{RQ2: Detecting Simple faults in Dependencies}
Our benchmark generated in total $1,122,420$ artificial updates for $311$ \maven
modules belonging to $262$ \github projects. \Cref{fig:regr:updates} shows a
violin plot of the mutation detection score for both direct and transitive
dependencies, split by project test suites on the left-hand side and \uppdatera
on the right-hand side. The median detection rate score is 51\% (mean: 47\%) for
direct dependencies and 36\% (mean: 35\%) for transitive dependencies. We can
observe that 25\% of the projects have a high test suite effectiveness greater
or equal to 80\% for direct dependencies. When looking at transitive
dependencies, the median of direct dependencies and the third-quantile of
transitive dependencies are similar, showing that only one-fourth of the
test suites remain effective in detecting faults in transitive dependencies.
Moreover, we can also see more dispersion in effectiveness among direct
dependencies than transitive dependencies, half of the projects have a detection
score raging between 16 to 54\% for transitive dependencies. Overall, the
results indicate that tests are effective for a limited number of cases and
dependencies. At large, however, a small minority of projects have test suites
that can comprehensively detect faulty updates. 

On the other hand, \uppdatera, has a median detection score of 97\% (mean: 74\%)
for direct dependencies and 88\% (mean: 64\%) for transitive dependencies.
Generally, we see that static analysis is highly effective in detecting simple
faults with a slightly decreased effectiveness for transitive dependencies. Half
of the projects with a low detection score ($< 50\%$) using tests now have
detection score greater than 80\%. In the lower half of the median for both
direct and transitive dependencies, we see large variations between the
projects. As change impact analysis is largely a generic technique, we manually
investigate why \uppdatera was unable to detect changes in 76 modules having a
low detection score of less than or equal to 39\% and 22\% for direct and
transitive dependencies, respectively. We perform a manual investigation using
the following protocol: (1) back-track from the dynamic call trace to test
suite, (2) identity potential tests cases that invoke the path in the call
trace, and (3) investigate both the test case setup and source code in-depth to
understand how \uppdatera could miss the regression change in the update.
\begin{sloppypar}
In total, we identified four potential reasons for \uppdatera to miss faulty
updates: 29 cases involving code generation, 26 cases involving class loading,
19 cases involving instrumentation, and 2 cases of instantiations of generic
methods. Dynamic class loading along with code generation makes use of Java's
Reflection API such as \texttt{Class.forName("DynClass");}. A majority of the
inspected cases stem from libraries such a FasterXML Jackson-databind, Jersey
REST framework, Spring framework, JAI ImageIO, Hibernate Validator, and Google
Guice. Reflection is useful in cases such as the creation of data bindings
(jackson-databind), data validation (hibernate or guice) or generation of HTTP
endpoints from annotated user methods (jersey or spring framework). Resolving
cases involving reflection is a known limitation of static
analysis~\cite{reif2019judge}.
\end{sloppypar}
Although we do not instrument \junit and \maven (which we use to power our
setup), projects can bypass our exclusion filter by putting those libraries
under a different namespace, a practice known as \textit{class shading}. We
identify several instances of bypassing the filter, an effect we cannot easily
control. Finally, in two cases, generic methods defined in user projects were
only instantiated in tests but not in the project source code. Generally, call
graph generators do not resolve generic methods unless there is a concrete
instantiation of it. 

\begin{table*}[tb]
\centering
\caption{Results of running \uppdatera on 22 \dependabot pull requests}
\label{tbl:manual:analysis}
\centering
\begin{adjustbox}{width={\textwidth},totalheight={\columnwidth},keepaspectratio}
\begin{tabular}{@{}lcccccrr@{}}
\toprule
\textbf{Pull Request} & \makecell{\textbf{Update} \\ \textbf{Type}} & \makecell{\textbf{Class}} &  \makecell{\textbf{Confirm}} & \makecell{\textbf{Test} \\ \textbf{Suite}} & \makecell{\textbf{\uppdatera}} & \makecell{\textbf{Test} \\ \textbf{Runtime }} & \makecell{\textbf{\uppdatera } \\ \textbf{Runtime}}\\
\midrule
\ahref{https://web.archive.org/web/20210129101235/https://github.com/spotify/dbeam/pull/189}{spotify/dbeam\#189}  & Patch & \textbf{S} & \cmark & FP   & FP    & 3.11 & 2.31     \\
\ahref{http://web.archive.org/web/20210129101154/https://github.com/airsonic/airsonic/pull/1622}{airsonic/airsonic\#1622} & Minor & \textbf{S}  &  \cmark      &  TP & \cellcolor[HTML]{FFFE65}{\color[HTML]{000000}FP}     & 77 & 7.5     \\
\ahref{http://web.archive.org/web/20210129101154/https://github.com/bitrich-info/xchange-stream/pull/570}{bitrich-info/xchange-stream\#570}    & Patch & \textbf{S} &  \cmark &  TP & \cellcolor[HTML]{FFFE65}{\color[HTML]{000000}FP}     & 2.78  & 1.93     \\
\ahref{http://web.archive.org/web/20201204024101/https://github.com/CROSSINGTUD/CryptoAnalysis/pull/245}{CROSSINGTUD/CryptoAnalysis\#245}    & Major & \textbf{S} &  \cmark   &  TP & \cellcolor[HTML]{FFFE65}{\color[HTML]{000000}FP}   & 12 & 2.6     \\
\ahref{http://web.archive.org/web/20200904205121/https://github.com/dbmdz/imageio-jnr/pull/84/}{dbmdz/imageio-jnr\#84}  & Patch & \textbf{S} &  \cmark & TP  & \cellcolor[HTML]{FFFE65}{\color[HTML]{000000}FP}   & -  & 0.7     \\
\ahref{http://web.archive.org/web/20210129101216/https://github.com/dnsimple/dnsimple-java/pull/23}{dnsimple/dnsimple-java\#23} & Minor & \textbf{S}    &  \cmark     &  TP & TP     & 2  & 11     \\
\ahref{http://web.archive.org/web/20210129101236/https://github.com/smallrye/smallrye-config/pull/289}{smallrye/smallrye-config\#289} & Patch & \textbf{S}   &  \cmark  &  TP & TP     & 1.1 & 0.6     \\
\ahref{http://web.archive.org/web/20200906190402/https://github.com/dropwizard/metrics/pull/1567/}{dropwizard/metrics\#1567}     & Patch & \textbf{S}   &  \cmark      &  TP & TP    & 2.6  & 8.73     \\
\ahref{http://web.archive.org/web/20210129101235/https://github.com/s4u/pgpverify-maven-plugin/pull/96}{s4u/pgpverify-maven-plugin\#96}    & Minor & \textbf{S}  &  \cmark     &  TP & TP      & 4 & 1.2     \\
\ahref{https://web.archive.org/web/20210129101255/https://github.com/JanusGraph/janusgraph/pull/2094}{JanusGraph/janusgraph\#2094}    & Minor & \textbf{N} &  \cmark   & FP & \cellcolor[HTML]{34CDF9}{\color[HTML]{000000}TN}    & 365 &33     \\
\ahref{http://web.archive.org/web/20210129101256/https://github.com/UniversalMediaServer/UniversalMediaServer/pull/1989}{UniversalMediaServer/UniversalMediaServer\#1989}   & Major & \textbf{U} &  \cmark &\cellcolor[HTML]{FE0000}{\color[HTML]{000000}FN} & TP     & 11 & 8.3     \\
\ahref{http://web.archive.org/web/20210104114503/https://github.com/premium-minds/pm-wicket-utils/pull/71}{premium-minds/pm-wicket-utils\#71}    & Patch & \textbf{N}  &  \cmark  & FP & \cellcolor[HTML]{34CDF9}{\color[HTML]{000000}TN}    & 1.56  & 0.51     \\
\ahref{https://web.archive.org/web/20210129101309/https://github.com/UniversalMediaServer/UniversalMediaServer/pull/1987}{UniversalMediaServer/UniversalMediaServer\#1987}     & Minor & \textbf{N}  &  \cmark  & FP & \cellcolor[HTML]{34CDF9}{\color[HTML]{000000}TN}    & 11 & 7.7     \\
\ahref{http://web.archive.org/web/20210129101308/https://github.com/CSUC/wos-times-cited-service/pull/36}{CSUC/wos-times-cited-service\#36}    & Patch & \textbf{S}  &  \xmark      &  TP & \cellcolor[HTML]{FFFE65}{\color[HTML]{000000}FP}     & 0.55 & 0.5     \\
\ahref{http://web.archive.org/web/20210129101313/https://github.com/Grundlefleck/ASM-NonClassloadingExtensions/pull/25}{Grundlefleck/ASM-NonClassloadingExtensions\#25}    & Major & \textbf{S} &  \xmark & TP & \cellcolor[HTML]{FFFE65}{\color[HTML]{000000}FP}    & 4 & 0.5     \\
\ahref{http://web.archive.org/web/20210129101329/https://github.com/RohanNagar/lightning/pull/211}{RohanNagar/lightning\#211}    & Major & \textbf{U}  &  \xmark  & TP & TP      & 2 & 7.8     \\
\ahref{http://web.archive.org/web/20201114175459/https://github.com/zalando/riptide/pull/932}{zalando/riptide\#932}    & Minor & \textbf{U} &  \xmark  & \cellcolor[HTML]{FE0000}{\color[HTML]{000000}FN}  & TP   & 7.5  & 20.5     \\
\ahref{http://web.archive.org/web/20210129101334/https://github.com/pinterest/secor/pull/1273}{pinterest/secor\#1273}   & Patch & \textbf{S}  &  \xmark & TP & TP   & 390 & 13.5     \\
\ahref{http://web.archive.org/web/20210129101335/https://github.com/michael-simons/neo4j-migrations/pull/60}{michael-simons/neo4j-migrations\#60}   & Patch & \textbf{S}  &  \xmark  &  TP & TP  & 3.45 & 0.8     \\
\ahref{http://web.archive.org/web/20210129101350/https://github.com/zaproxy/crawljax/pull/115}{zaproxy/crawljax\#115}     & Minor & \textbf{U}  &  \xmark &  \cellcolor[HTML]{FE0000}{\color[HTML]{000000}FN} & TP   & 17.35 & 1.3     \\
\ahref{http://web.archive.org/web/20210129101356/https://github.com/hub4j/github-api/pull/793}{hub4j/github-api\#793}     & Minor & \textbf{S}   &  \xmark  &  TP & TP   & 1.3 & 4    \\
\ahref{http://web.archive.org/web/20210129101401/https://github.com/zalando/logbook/pull/750}{zalando/logbook\#750}     & Patch & \textbf{S}   &  \xmark  & TP & TP & 6.1 & 18.38     \\

\bottomrule
\end{tabular}
\end{adjustbox}
\label{table:performance}
\end{table*}

\begin{mdframed}[roundcorner=5pt,nobreak=true,align=center]
 \textbf{Findings from RQ2:}~\textit{
Project tests are effective in a limited number of cases but not at large.
\uppdatera can detect twice as many faulty artificial updates as opposed to project
test suites. Libraries making use of Java's reflection API could affect its
applicability.}
\end{mdframed}

\subsection{RQ3: Change Impact Analysis in Practice}
We conducted our online monitoring for two weeks between 13-27 Apr 2020
evaluating in total 22 \dependabot pull requests. On average, we harvested
around 350 pull requests per day between Mondays and Wednesdays, 150 pull
requests per day between Thursday and Fridays, 50 pull requests per day on the
weekends. While the number of pull requests may seem high, a majority of them
were updates of \maven plugins or test dependencies, uncompilable, or
superseding previous pull requests. Thus, we posted on average two pull
requests per day taking anywhere between one to four hours to manually evaluate
pull requests and post our findings as comments.

\begin{sloppypar}
\Cref{tbl:manual:analysis} presents the analyzed pull requests along with the
update type, ground truth class~(i.e., \textbf{Class} column), external
confirmation~(i.e., \textbf{Confirm} column), results from the tooling, and
execution times (in minutes). In total, our ground truth consists of 15 pull
requests where the update is safe (i.e., \textbf{S} class), three pull requests
where the dependency under update is unused and only declared (i.e., \textbf{N}
class), and four pull requests where the updates that are unsafe (i.e., \textbf{U}
class).  The test suites of the analyzed pull requests classified 15 update
as true positives(TP), four update cases as false positives (FP), and three
update cases as false negatives (FN). \uppdatera classified 12 update cases as
true positives (TP), seven cases as false positives (FP), three cases as true
negatives (TN). There are 12 cases where the two techniques report differently
as highlighted by the colors in \Cref{tbl:manual:analysis}. Most notable are
false positives; \uppdatera incorrectly reports six updates (highlighted
yellow in the table) as unsafe that test suites can detect as safe. In those
cases, the heuristics failed to account for refactorings or falsely derived call
paths due to dynamic dispatch. In four cases,~\uppdatera could not detect that
the changes were refactorings (i.e., semantic-preserving changes). One such
example is a confirmed minor update of the Apache \texttt{commons-lang3} library
refactoring array length and null checks into a new function. In the two
remaining cases, all reachable call paths were over-approximations. The update
of \texttt{org.eclipse.emf.common} in one project included changes to List
structures implementing methods of Java's List Interface (such as
\texttt{addAll()}), resulting in unrelated interface calls being linked to it.
This is a limitation of the CHA algorithm as it links interface calls to all
available implementations. In three confirmed \textbf{N}-cases (highlighted blue in the
table) where tests would falsely pass the updates, \uppdatera correctly
identified no use of the dependency under update in the projects. The project
maintainers in two of the reported cases have started refactoring work to remove
those identified dependencies.
\end{sloppypar}
\begin{sloppypar}
\uppdatera was able to complement test suites in three false-negative cases
(highlighted red in the table). In our confirmed case of an unsafe update,
\uppdatera identified the Apache \texttt{commons-lang3}
library to break the application logic of a project due to changes in
calculating string edits using the Jaro–Winkler distance. Generally, we can
observe that solely using static analysis may risk falsely classifying safe
updates as unsafe. Finally, we also make a comparison of execution times between
running tests and \uppdatera. The results reveals that \uppdatera has faster or
comparable times in 16 out of 22 cases, suggesting that change impact analysis
can be a viable option to complement tests in CI environments.
\end{sloppypar}
\begin{mdframed}[roundcorner=5pt,nobreak=true,align=center]
\textbf{Findings from RQ3:}~\textit{Semantically equivalent changes
(refactorings) and over-approximated function calls are the main sources of
false positives in \uppdatera. However, \uppdatera helped project maintainers
identify risky updates and unused dependencies.}
\end{mdframed}

\section{Discussion}

\subsection{Evaluating Library Updates}
Updating to a new version of a third-party library is not a
trivial task, and for good reasons: interface refactorings induce additional
maintenance burden and integrating untested behavior can jeopardize project
stability. Services such as \dependabot advocate a modest update strategy
focusing on project compatibility: only update if the tests pass with the new
library version. Effectively, developer-written tests act as the first-line
defense against library updates introducing regression changes.

A key insight in our work is that automated dependency updates are not reliable.
Our results strongly suggest that existing developer-written tests lack
specifications that exercise dependencies in depth. This finding is in line with
the work by Mirhosseini et al.~\cite{mirhosseini2017can}, where developers
report being suspicious of integrating automated updates due to fear of
breakage. When selecting to adopt a third-party library, Bogart et al., report
that developers look at aspects such as reputation, code quality standards and
active maintenance to build up trust~\cite{bogart2016break}. Perceived
high-quality libraries can eliminate the need for extensive testing. In our
case, we found evidence against this practice. The minor backward-compatible
update of \texttt{org.apache.commons:commons-lang3}, a
high-quality library, had changes that would break the application logic in one
project if the pull request was merged in our manual analysis. In addition, the practice of
testing third-party libraries is not common among popular testing
books~\cite{whittaker2002break,myers2011art,hetzel1988complete}, very few
research papers suggest testing of third-party
libraries~\cite{kropp1998automated,mariani2007compatibility}.

Directing testing efforts to dependencies would be a potential solution to the
problem. Therefore, we recommend practitioners to use automated updating
services cautiously and complement with tests for critical library dependencies.
For tool creators in the domain, we argue for increased transparency in
automated updating. With a small minority of projects having both coverage and
tests capable of detecting simple regressions, pull requests could feature a
confidence score on how well it is able to test new changes in a library under
update. As a first step, tool creators can make use of our study setup to
measure both coverage and quality of tests as an indication of confidence. A
confidence score could also help reduce false negatives: if no tests are
exercising a changed functionality of a dependency under update, \dependabot
could avoid recommending it.

\subsection{Strengths and Weaknesses of Static Analysis}
Without needing to maintain additional dependency-specific tests, static
analysis can be effective in deterring updates with potential regression
changes. For a large number of projects with limited test quality, change impact
analysis can fill the gap where tests are unable to reach and would be a
compelling option for tool creators to consider. For a minority of projects,
however, we identify certain third-party libraries that impede the overall
analysis accuracy. Libraries heavily relying on code generation such as the
Spring framework makes use of the Java Reflection API that are known to be
statically difficult to analyze~\cite{antoniadisstatic}, could miss critical
execution paths in projects that make use of them. Moreover, by linking
interface calls to all its implementations, call graphs contain
over-approximated call paths. We could observe non-existing interface calls from
functions in the unused dependency to classes implementing the interface in the
project during the manual analysis. As Ponta et al.~\cite{ponta2018beyond}
approach base on building a call graph with the project and its dependencies
together, we make preliminary observations that projects having library
dependencies with several common interfaces between them are likely to have many
unrelated function calls. Exploring improvements such as using type hints with
data flow analysis could potentially eliminate such function calls. Overall, we
argue that static analysis is a useful complement in use cases where tests lack
coverage. By also revealing and presenting gaps and quality issues in test
suites, static analysis can help developers in prioritizing testing efforts of
dependencies. 

\subsection{Threats to Validity}\label{sec:threats}

Sampling random projects from \github pose threats to our results: tests or
dependencies in projects may not exercise production classes. To mitigate this
risk, we configure our call extractor to only record call paths originating from
the project source code. Call paths that do not traverse via project source code
are excluded (e.g., test class directly calling a dependency).
\begin{sloppypar}
The use of mutation analysis to emulate source code changes in dependency
functions has several potential threats to validity. First, we acknowledge that
the applied mutation operators do not substitute actual regression changes in
library updates. Our objective is to exercise all uses of libraries in a project
by injecting simple faults to uncover potential coverage gaps in updating tools.
Using real-world cases for this purpose would be challenging and potentially
adding hidden uncontrolled factors. Second, our ground truth in \textbf{RQ2}
represents reachable call paths inferred from running project tests, making it a
subset of all possible executions and is a limitation of the benchmark. A
potential avenue to explore is the use of test generation techniques such as
EvoSuite~\cite{fraser2011evosuite} to discover new call paths. However, EvoSuite
generates tests at the class level without considering its interaction with
other classes or dependencies, generating artificial tests that may not represent
valid use cases.
\end{sloppypar}
The false-positive rate in \textbf{RQ3} is indicative and not representative.
Without domain knowledge of the interplay between a project and a dependency,
the code reviews may state incorrect or incomplete information. To mitigate this
risk, we post our code review assessment in the update for the project
maintainer to react in case of incorrect analysis. Finally, for the
reproducibility of our study, we have made the source
code,\footnote{\url{https://github.com/jhejderup/uppdatera}} the experimental
pipeline,\footnote{\url{https://github.com/jhejderup/uppdatera-pipeline}} and
our data publicly available~\cite{uppdatera:ds}. Specifically, we include the
examined projects, applied mutation changes, and their dynamic and static call
graph.

\section{Related Work}
\paragraph{{\normalfont \textbf{Updating library dependencies in projects}}}
To assist developers with updating dependencies in projects, researchers have
studied practices around updating
dependencies~\cite{kula2018developers,mirhosseini2017can,bogart2016break,dietrich2019dependency,dietrich2014broken,raemaekers2017semantic}
and proposed tools leveraging both static- and dynamic
analysis~\cite{foo2018efficient,mezzetti2018type,moller2019model}. Kula
\textit{et al}.~\cite{kula2018developers} empirical study of $2,700$ library
dependencies in $4,600$ Java project found that $81.5$\% remain outdated, even
with security problems. The study found that factors such as uncertainties
around estimating refactoring efforts and other task priorities as reasons for
developers to not update dependencies. To address the update fatigue for
developers, automated dependency updaters such as \dependabot
and \texttt{greenkeeper.io} actively reminds and suggests dependency updates to
developers through the use of pull requests. A study by Mirhosseini \textit{et
al}.~\cite{mirhosseini2017can} found that pull requests encourage developers to
update dependencies more frequently but the frequency of updates and lack of
convincing arguments defer them from updating. On similar lines, the work of
Bogart \textit{et al}.~\cite{bogart2016break} also suggests that developers
perceive the use of monitoring tools to have a high signal-to-noise ratio than
giving actionable insights. Finally, the empirical work of Dietrich~\textit{et
al.}~\cite{dietrich2014broken} suggests that 75\% of emulated library updates in
the Qualitas dataset has breaking changes. However, only a few updates resulted
in an error, motivating the need for contextual analysis.

Recently researchers have started to explore the use of static- and dynamic
analysis to identify library updates with breaking changes, saving developers
time, and review efforts of library updates.
\texttt{NoRegrets}~\cite{mezzetti2018type,moller2019model} is a tool that
detects breaking changes in test suites of dependent \npm packages before
releasing an update of the library. Although helpful in minimizing the chances
of breaking changes for clients, the identified subset of clients may not be
representative of other clients. Similarly, Foo \textit{et
al.}~\cite{foo2018efficient} describes a static approach using simple diffing
and querying Veracode's SGL~\cite{foo2018sgl} graph to find clients affected by
breaking changes. In contrast to this approach, \uppdatera analyzes at the project level
(e.g., does not search for affected clients), targets diff with data- and
control flow changes (i.e., not only interface changes), and includes a
benchmark to compare updating tools.

\paragraph{{\normalfont \textbf{Change Impact Analysis}}}
Change Impact analysis is a widely studied problem in program analysis
research~\cite{li2013survey,lehnert2011taxonomy}. Propagation of changes in
package repositories have become an important research area in light of incidents
such as the \textit{left-pad incident}, and recent moves to emulate these
problems on package-based
networks~\cite{abdalkareem2017developers,kikas2017structure}. Several
techniques~\cite{ryder2001change,badri2005supporting,ren2004chianti,german2009change,li2012combining}
use call graphs as an intermediate representation for change impact analysis.
Alternative techniques to call graphs are static and dynamic
slicing~\cite{tip1994survey, arnold1996software},
profiling~\cite{law2003whole,orso2003leveraging} and execution
traces~\cite{orso2004empirical}. Due to cost-precision trade-offs, several
proposed approaches use a combination of these techniques. One such example is
Alimadadi~\textit{et al}.'s work on Tochal, that leverages both runtime data and
call graphs to more accurately represent changes to dynamic features such as the
DOM. For a comprehensive overview of impact analysis techniques and change
estimations, we refer the reader to Li~\textit{et al}'s~\cite{li2013survey}
survey on code-based change impact analysis techniques

An application of change impact analysis is regression test
selection techniques~\cite{yoo2012regression} (RTS) such as class-based
STARTS~\cite{shi2019reflection,legunsen2017starts} and probabilistic test
selection~\cite{machalica2019predictive} that find relevant tests for evaluating
new code changes. We found in our evaluation that test suites have limited
coverage of dependencies, thus RTS may not be able to find tests relevant for
changes in dependencies or have enough test data to build a prediction model for
average \github projects. Finally, Danglot~\textit{et
al}.~\cite{danglot2020approach} and Da Silva~\textit{et
al}.~\cite{da2020detecting} investigate the use of search-based methods such as
test amplification and automated test generation for detecting semantically
conflicting changes. Although search-based methods are effective in reducing
false positives and to some degree eliminating false negatives present in static
analysis, they are limiting for integration test scenarios such as automated
dependency updating. Da Silva~\textit{et al}.~\cite{da2020detecting} found that
automated test generation such as EvoSuite~\cite{fraser2011evosuite} have
difficulties in generating effective tests for complex objects with internal or
external dependencies.

\section{Conclusions and future work}
In this paper, we empirically investigate the reliability of test suites for
automating dependency updates. With an increasing number of developers relying
on services that automate updates of dependencies, our goal was to uncover to
what degree project tests exercise utilized functionality in library
dependencies, how effective they are in catching simple regressions, and there
performance in practice. As recent research highlights the need for conservative
techniques, we explored the use of change impact analysis to reduce false
negatives.

Our findings show that half of 521 well-tested projects with tests cover less
than 60\% of their function calls to direct dependencies. The coverage drops to
20\% when considering call paths to transitive dependencies. By artificially
injecting simple faults in library dependencies to 262 projects, we observe that
one-forth of the projects can detect 80\% or more faults in functions of direct
libraries. When considering transitive dependencies, the number of projects drop
to one-eighth. Conversely change impact analysis, can detect 80\% of potentially
breaking in changes in both direct and transitive dependencies, two times more
than using test suites. Although change impact analysis is a promising direction
to flag faulty updates, we also manually investigate whether it can complement
tests in 22 \dependabot pull requests. Our results show that change impact
analysis could avoid unsafe updates in three cases where tests failed and
spotted unused libraries in two cases. However, there are more false positives
with change impact analysis as it is more imprecise than tests.

Our findings suggest that developers that are making use of automated dependency
updating need to be aware of the risks with using project tests for
compatibility checking.  Without coverage or adequate tests for all usages of
library dependencies, updates can silently introduce unintended functionality
over time. As services such as \dependabot do not advertise risks involved with
updating dependencies, tool creators could introduce reliability measurements
such as scoring test suites in pull requests. As we investigate the use of
change impact analysis, we argue that tool creators should explore combining
dynamic and static analysis to derive verification techniques that do not
strongly depend on users' test suites. 

\begin{sloppypar}
In future work, we aim to establish best practices for updating
third-party libraries. As a first step, we aim to understand whether
developers direct testing efforts towards dependencies and uncover the strategies they use.
Moreover, we also intend to explore hybrid workflows through data-driven methods
for efficient update checking by combining dynamic and static
analysis.
\end{sloppypar}

\section*{Acknowledgment}
We thank Moritz, Xunhui, Arie, Mauricio, and Ayushi for reviewing drafts of this
paper. The work in this paper was partially funded by NWO grant 628.008.001
(CodeFeedr) and H2020 grant 825328 (FASTEN). 

\bibliographystyle{elsarticle-num}
\bibliography{semdate}

\begin{wrapfigure}{l}{25mm} 
    \includegraphics[width=1in,height=1.25in,clip,keepaspectratio]{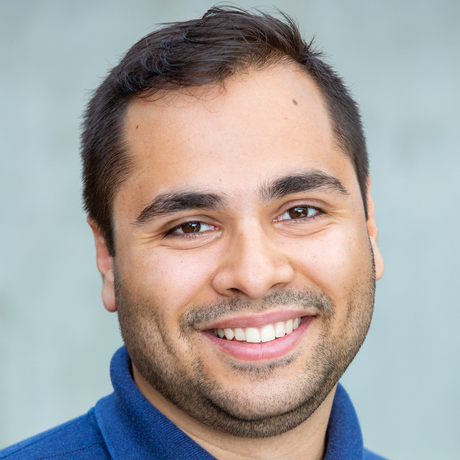}
  \end{wrapfigure}\par
  \noindent\textbf{Joseph Hejderup} is a Ph.D. student at the Delft University of Technology, the Netherlands. His primary research interest is to make package management systems more intelligent, safe, and robust using program analysis and empirical methods. He is the main author of Präzi, a technique that constructs fine-grained dependency networks using call graphs. He holds an M.Sc. from the Delft University of Technology, the Netherlands.\\\\\par

\begin{wrapfigure}{l}{25mm} 
\includegraphics[width=1in,height=1.25in,clip,keepaspectratio]{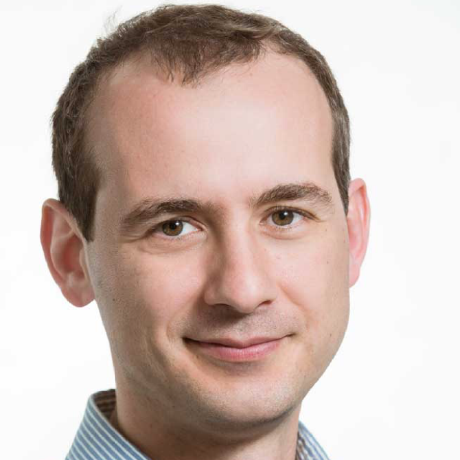}
\end{wrapfigure}\par
\noindent\textbf{Georgios Gousios} 
 is a research engineer at Facebook and an associate professor at the Delft University of Technology, 
 The Netherlands (on leave). He works in the fields of software analytics, software ecosystems, 
 software processes, and machine learning for software engineering. 
 He is the main author of the GHTorrent data collection and curation framework 
 and various widely used tools and datasets.
\end{document}

%% file: abbreviations.tex
\usepackage{xspace}

\newcommand{\npm}{\textsc{npm}\xspace}
\newcommand\maven{\textsc{Maven}\xspace}

\newcommand\junit{\textsc{JUnit}\xspace}
\newcommand\testng{\textsc{TestNG}\xspace}

\newcommand\dependabot{\textsc{Dependabot}\xspace}
\newcommand\uppdatera{\textsc{Uppdatera}\xspace}
\newcommand\crates{\textsc{Crates.io}\xspace}
\newcommand\mavencentral{\textsc{Maven Central}\xspace}

\newcommand\cargo{\textsc{Cargo}\xspace}

\newcommand\travis{\textsc{Travis CI}\xspace}

\newcommand\github{\textsc{GitHub}\xspace}
\newcommand\ghtorrent{\textsc{GHTorrent}\xspace}